# Physica Scripta

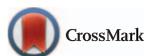

**PERSPECTIVE**

OPEN ACCESS

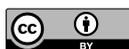

# The sounds of science—a symphony for many instruments and voices: part II

Gerard 't Hooft[1], William D Phillips[2], Anton Zeilinger[3,4], Roland Allen[5], Jim Baggott[6], François R Bouchet[7], Solange M G Cantanhede[8], Lázaro A M Castanedo[9,10], Ana María Cetto[11], Alan A Coley[12], Bryan J Dalton[13], Peyman Fahimi[9,14], Sharon Franks[15], Alex Frano[15], Edward S Fry[5], Steven Goldfarb[16,17], Karlheinz Langanke[18,19], Chérif F Matta[9,10,14,20], Dimitri Nanopoulos[5], Chad Orzel[21], Sam Patrick[22,23], Viraj A A Sanghai[12], Ivan K Schuller[15], Oleg Shpyrko[15] and Suzy Lidström[5,*]

[1] Utrecht University, Institute for Theoretical Physics, Postbox 80.089, Utrecht, 3508 TB The Netherlands
[2] Joint Quantum Institute, University of Maryland, National Institute of Standards and Technology, Gaithersburg, 20899-8424, MD, United States of America
[3] Vienna Center for Quantum Science & Technology (VCQ), Faculty of Physics, University of Vienna, Boltzmanngasse 5, Vienna, 1090 Austria
[4] Institute for Quantum Optics and Quantum Information (IQOQI), Austrian Academy of Sciences, Boltzmanngasse 3, Vienna, 1090, Austria
[5] Department of Physics and Astronomy, Texas A&M University, College Station, TX, United States of America
[6] Freelance science writer based in Cape Town, South Africa
[7] CNRS; Université de Sorbonne, Institut d'astrophysique de Paris, 98 Bis Boulevard Arago F-75014, Paris, France
[8] Psychoanalyst, Member of the Brazilian Psychoanalysis Space (EBEP), Rio de Janeiro, Brazil
[9] Department of Chemistry and Physics, Mount Saint Vincent University, Halifax, Nova Scotia, B3M2J6, Canada
[10] Department of Chemistry, Saint Mary's University, Halifax, Nova Scotia, B3H 3C3, Canada
[11] Instituto de Física, Universidad Nacional Autónoma de México, Ciudad Universitaria, Mexico City, Mexico
[12] Department of Mathematics and Statistics, Dalhousie University, Halifax, B3H 4R2, Nova Scotia, Canada
[13] Centre for Quantum Science and Technology Theory, Swinburne University of Technology, Melbourne, 3122, Victoria, Australia
[14] Départment de chimie, Université Laval, Québec, Québec, G1V 086, Canada
[15] University of California-San Diego, La Jolla, CA, 92093, United States of America
[16] School of Physics, The University of Melbourne, Grattan Street, Parkville, Victoria, 3010 Australia
[17] ATLAS Experiment, CERN, Geneva, Switzerland
[18] GSI Helmholtzzentrum für Schwerionenforschung, Darmstadt, Germany
[19] Institut für Theoretische Physik, Technische Universität Darmstadt, Darmstadt, Germany
[20] Department of Chemistry, Dalhousie University, Halifax, Nova Scotia, B3H 4J3, Canada
[21] Physics and Astronomy Department, Union College, 807 Union Street, Schenectady, 12308, NY, United States of America
[22] University of British Columbia, Vancouver, V6T 2A6, Canada
[23] Institute for Quantum Science and Engineering, Texas A & M University, College Station, TX, United States of America
* Author to whom any correspondence should be addressed.

E-mail: g.thooft@uu.nl, william.phillips@nist.gov, anton.zeilinger@univie.ac.at, allen@tamu.edu, jim@logosconsulting.co.uk, bouchet@iap.fr, solangecam@gmail.com, lamonteserincastanedo@gmail.com, ana@fisica.unam.mx, aac@mathstat.dal.ca, peyman.fahimi.1@ulaval.ca, sfranks@ucsd.edu, afrano@ucsd.edu, fry@physics.tamu.edu, Steven.Goldfarb@cern.ch, k.langanke@gsi.de, Cherif.Matta@msvu.ca, dimitri@physics.tamu.edu, orzelc@union.edu, sampatrick31@googlemail.com, oshpyrko@physics.ucsd.edu and suzy.lidstrom@gmail.com

**Abstract**
Despite its amazing quantitative successes and contributions to revolutionary technologies, physics currently faces many unsolved mysteries ranging from the meaning of quantum mechanics to the nature of the dark energy that will determine the future of the Universe. It is clearly prohibitive for the general reader, and even the best informed physicists, to follow the vast number of technical papers published in the thousands of specialized journals. For this reason, we have asked the leading experts across many of the most important areas of physics to summarise their global assessment of some of the most important issues. In lieu of an extremely long abstract summarising the contents, we invite the reader to look at the section headings and their authors, and then to indulge in a feast of stimulating topics spanning the current frontiers of fundamental physics from 'The Future of Physics' by William D Phillips and 'What characterises topological effects in physics?' by Gerard 't Hooft through the contributions of the widest imaginable range of world leaders in their respective areas. This paper is presented as a preface to exciting developments by senior and young scientists in the years that lie ahead, and a complement to the less authoritative popular accounts by journalists.

## 1. Prelude to the second movement by Suzy Lidström

The first movement of *Sounds of Science—A symphony for Many Instruments and Voices* [1] terminated with the words 'to be continued' and an ellipsis, emphasising the authors' understanding that Nature is a wonderous mystery well worth probing. This is a view shared by Sean Carroll, who, reflecting on Albert Camus' belief that the Universe was 'unintelligible', countered: 'It is actually the opposite of that—the fact that the Universe is so gloriously knowable is perhaps its most remarkable feature' [2].

Dr. Zdeněk Papoušek's words to the participants of FQMT—*Frontiers of Quantum and Mesoscopic Thermodynamics*—held in Prague, were still resonating when the interlude commenced. He encouraged the audience, informing us that:

> There is a scientist, a philosopher and an artist in every one of us. There are certain things that we need to test; we need to test them all and then hold on to what is good. Other things, though, shall make us wonder and ask questions, even without getting the answers.
>
> *Zdeněk Papoušek, Chairman of the Committee on Education, Science, Culture, Human Rights and Petition of the Senate of the Czech Republic*

In this spirit, the second movement of *Sounds of Science—A Symphony for Many Instruments and Voices* continues our questions-based reflections, in particular, presenting variations on a theme: *Will there be new physics?* This question was proposed for discussion by the scientific community by a young researcher attending the aforementioned conference. He explained that his intention was to stimulate reflection on whether further paradigm shifts of the magnitude of that represented by the transition from classical to quantum worldviews might be anticipated in the future. We have, however, deliberately sought to broaden the interpretation of his question in seeking responses, thereby reflecting the Czech cell biologist's belief that:

> Science should strike as many sparks as one's sight can bear.
>
> *Jan Evangelista Purkyne*

Some themes from the first movement [1] reverberate here alongside new ones, collectively adding to our earlier work, composed with the interested scientist in mind [3–5], and complementing publications by prominent scientists who have written for the general public [2, 6, 7]. Although Stephen Hawking passed away before seeing his final oeuvre *Brief Answers to the Big Questions* through to publication [7], a final message was broadcast posthumously inviting everyone to 'look up at the stars and not down at [y]our feet'. Hawking's voice encouraged people to contemplate the benefits of science and technology, and to: 'Try to make sense of what you see, and wonder about what makes the Universe exist'. In this spirit, rather than present novel research results, in our 'Perspective' paper we contemplate the future of our respective fields.

Standard physics at its most fundamental level is now entirely described by quantum fields, and this description has proved quantitatively accurate to about ten significant figures. However one can imagine the potential for a deeper description to give rise to quantum fields in an effective theory at the energy scales that are now accessible to experiment. In many contributions, authors address this issue from various perspectives with some emphasis on the prominent mysteries that seem to point to new physics, while we keep in mind that the final arbiter will be experiment. Thus, we pass freely between theory and experiment as we consider different areas of interest.

The paper opens with *The Future of Physics*, in which William Phillips (figure 1) places present research in its historical context, preparing the way for subsequent authors to provide a perspective from within their own specialist areas. Gerhard 't Hooft (figure 1) subsequently poses the question: *What characterises topological effects in physics?*, revealing the fascination of this topic. The first of several contributions bearing the title: *Will there be new physics?* follows when Dimitri Nanopoulos addresses some areas of fundamental physics under the subtitle: *From Classical -> Quantum -> ?*

The diversity of the issues to be tackled at the new international experimental Facility for Antiproton and Ion Research (FAIR) in Darmstadt is evident in Karlheinz Langanke's *FAIR—Exploring the Universe in the laboratory*. In the subsequent piece, Edward Fry connects observation with theory as he discusses the challenge of comprehending the nature of reality as we experience it, and relating this experience to quantum phenomena, asking: *How does a quantum measurement decide which outcome is observed?*

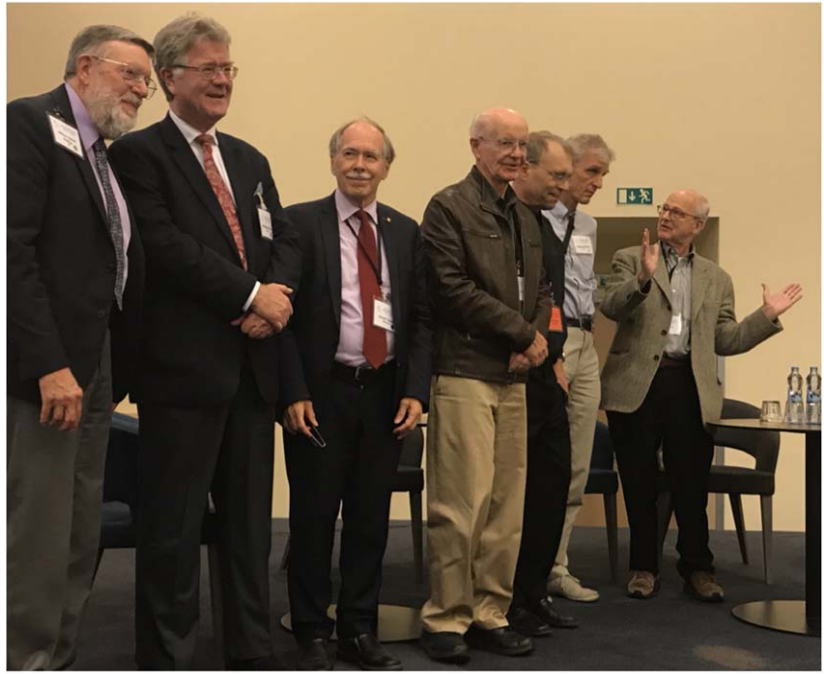

**Figure 1.** On the stage at FQMT in Prague (from left to right): William D. Phillips, Nobel laureate for his work on laser cooling; Wolfgang Schleich, Acting Director of the German Aerospace Center's DLR Institute of Quantum Technologies and of the Institut für Quantenphysik; Gerard 't Hooft, Nobel laureate for elucidating the quantum structure of electroweak interactions; Marlan Scully, Director of the Institute for Quantum Science and Engineering (IQSE) and the Center for Theoretical Physics; Vaclav Spicka organiser of FQMT and the magnificent series of concerts associated with it; Wolfgang Ketterle, Nobel laureate for his work on Bose-Einstein condensation; and Rainer Weiss, Nobel laureate for the introduction of gravitational wave astronomy. Professors Schleich, Scully, and Weiss contributed to previous papers of this kind [1, 5]. Photograph: Suzy Lidström.

Returning to the principal question, *Will there be new physics?*, in *Is there new physics beyond the Standard Model?*, François Bouchet reminds us that it is not merely aesthetics that suggests that the Standard Model of particle physics should be supplanted by a more complete theory: theory, experiment and observation essentially compel us to find a more satisfying vision of Nature. Chad Orzel offers a broad perspective on the current challenges, concluding that: *We're not done with the old physics yet*. In another piece bearing a subtitle, *Hello darkness my old friend*, Alan Coley and Viraj Sangai discuss an astounding mystery touched on by William Phillips in his introduction, asking: *What is the dark energy in cosmology?* Sam Patrick introduces the role of analogue gravity in: *Are the secrets of the Universe hiding in your bathtub?* The subsequent piece by Jim Baggott presents his thoughts on the question: *Will there be new physics?* with an emphasis on research that he believes does not merit funding. Opinions often differ in scientific discourse, and a healthy dialogue reflecting contrasting views is a vital part of the process towards the truth.

In the subsequent piece, Roland Allen takes his readers on a journey to consider the glorious variety of multiverses, tackling the question: *How big is Nature, and how much of it can we explore?* Ivan Schuller and his colleagues are engaged on a journey of exploration of a very different kind, one that is: *Towards a machine that works like a brain*. Chérif F. Matta and his coworkers lead us to another new realm, to consider the question: *What can we say about the 'value of information' in biophysics?* Uniting the previous themes of biophysics and brains, Suzy Lidström and Solange Cantanhede examine what we know of how consciousness emerges in individuals under a title inspired by Stephen Hawking's famous question [8]: *What breathes the fire of consciousness into our brains?*

*What philosophers should really be thinking about* by Roland Allen and Suzy Lidström follows. In *How can scientists address misinformation?* Steven Goldfarb brings his experience as a science communicator to bear as he seeks to convince researchers of the need to redouble their efforts at outreach to address misinformation and encourage fact-based decision-making by world leaders.

A technical piece follows in which Bryan Dalton asks: *Can we find violations of Bell locality in macroscopic systems?* Then the tempo changes again as Ana Maria Cetto, shown enthusing young visitors to the Museum of Light at the Autonomous National University of Mexico (UNAM) in (figure 2), presents her chosen topic: *What is the source of quantum non-locality?*

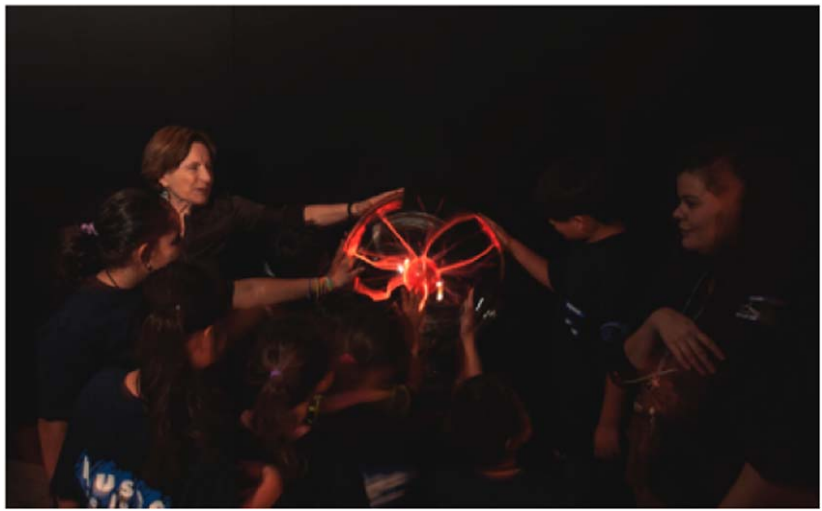

**Figure 2.** Visitors to the Museum of Light (UNAM) admiring a 'lightning strike' within a plasma sphere. The Museum's Director Ana Maria Cetto (left) explains the physics to an enthusiastic young group of onlookers. Credit: Arturo Orta.

As the end of the article approaches, Anton Zeilinger provides a broad perspective reflecting on progress made under the title: *How much of physics have we found so far?*

The instruments and voices reach far beyond the quantum and mesoscopic themes of the conference at which the majority of questions were gathered. In doing so, they explore themes, arriving at different interpretations. An openness to discourse should be welcomed in the scientific community, with experimental results and observations being the ultimate arbiter, as mentioned earlier.

We hope you will enjoy the performance.

## 2. The future of physics by William D Phillips

According to an oft-repeated legend, near the 1900 turn of the century, physicists held the opinion that they understood everything pretty well and all that was left in physics was to add more decimal places to the measured numbers characterizing the physical world. Regardless of the truth of that legend (and surely it was true for at least some well-known physicists) nothing could have been further from the truth. We were about to embark on what was arguably the most revolutionary period in the development of physics.

The dawn of the 20th century saw Max Planck explain the spectrum of thermal radiation by assuming that energy is exchanged between radiation and matter in discrete packages or quanta. This is often seen as being the beginning of quantum mechanics, the greatest scientific, technological, and philosophical revolution of the century. In fact, a clearer beginning of quantum mechanics was in Einstein's explanation of the photoelectric effect, one of the fruits of his 1905 *annus mirabilis*, in which he proposed that light is actually composed of packets of energy, which we now call photons. Further key insights by people like Bohr, Heisenberg, Schroedinger, and Dirac produced a well-developed quantum theory by about 1930.

Returning to Einstein and his miraculous year, we find two more revolutionary works: special relativity, which changed our very notions of space and time, and Brownian motion, which finally cemented the understanding that matter is made of atoms and molecules—a concept still widely resisted before Einstein. A decade later, Einstein's theory of general relativity had upended our understanding of gravity, and with it, even more deeply revolutionized the ideas of space and time, now seen as a unified fabric of the Universe.

So, not long after the predictions that physics was over, we had embarked on an adventure that took physics into totally unanticipated directions defined by atomic theory, quantum mechanics, and relativity.

If the turn of the 20th century saw such wrong-headed ideas about the future (or, the lack of a future) for physics, what about our own century? Around the year 2000, a number of popular scientific books proclaimed *The End of Physics* [9] or *The End of Science* [10], positing that we had already discovered all there was to know, and what remained unknown was so difficult and beyond our ability to explore that we would never know it. I remember attending a seminar by one of these prophets of stagnation who ended his talk with a consolation to the physicists, who would no longer experience the joys of discovery, by reminding us, tongue in cheek, 'There is still sex and beer.' The physicists were not buying it.

As I see it, we live in an incredibly exciting time for physics in particular and for science in general. We now know, with a reasonable degree of precision, that about 5% of the mass-energy of the Universe is made up of stuff we understand: hydrogen and other elements, or constituents like protons, neutrons, quarks, electrons, muons, neutrinos, photons, and the other fundamental particles of the Standard Model. Five percent! The rest is about 25% dark matter, about which we understand nothing, and about 70% dark energy, about which we understand even less. What could be more exciting than to inhabit a universe where about 95% of everything is waiting to be understood? We know that two of the most well-established theories ever devised—General Relativity and Quantum Mechanics—theories whose tight construction is pure beauty, are incompatible with each other. There is another theory, waiting to be discovered, that will unify these two. These are only a sampling of what we do not yet understand. And to make matters even more delicious, experiments are underway that may provide clues to the solution of these mysteries in my lifetime. The full solution will probably take longer, but considering that it was a few centuries between Newton and Einstein, that is no surprise. Truly fundamental changes in our understanding of physics await, and I am eager to see some of those changes and perhaps even participate in them.

But such fundamental new discoveries, which I am confident will come, are not the only reason that the turn-of-the-century naysayers were so deeply mistaken. In my own field of research, experimental atomic, molecular, and optical (AMO) physics, we have understood the needed fundamentals since about 1930. Yet, as an AMO community, we are surprised every day by things we learn in the laboratory, and enlightened every day by the new insights of our theoretically inclined colleagues. And that same scenario plays out in the other subfields of physics. Furthermore, the insights and techniques of today's physics are being applied to chemistry and biology, opening revolutionary, wholly unanticipated, exciting research directions in those fields. No, physics is in no danger of coming to an end in our lifetimes, or in the lifetimes of our great-great-grandchildren. I have confidence that the great intellectual adventure of understanding the inner workings of nature will never come to an end. Each new discovery produces not just understanding, but new questions. Each new technology makes possible new fundamental discoveries that lead to new technologies. The unending ingenuity of the human spirit ensures that science will always be an endless frontier.

## 3. What characterises topological effects in physics? by Gerard 't Hooft

One may question what it means to call some physical phenomenon 'topological'. In practice, one constructs mathematical models, and in these models one can sometimes recognise typically geometrical considerations to classify structures that could be particles, events or more extended, non-local features. But there are also many mysteries in the physical world that we have not yet managed to frame in a model. Certain characteristics then make one suspect that these will also hang together with general geometrical structures that are independent of dynamical, mechanical details.

There are numerous phenomena in the world of physics that can be understood as effects of a topological nature. Often, these are features that come as surprises. A famous example is the *soliton*. A soliton is a solution of some dynamical wave equation that behaves as a particle, instead of spreading out and disappearing. It looks as if there is something that prevents the solution from behaving as ordinary waves. A typical example is a strong wave crest travelling in a channel, so that it looks like a particle in one dimension, but also tsunamis behave somewhat like a soliton, travelling thousands of miles without any tendency to spread out.

A soliton solution carries mass, energy and momentum, and indeed, it resembles a particle so much that investigators began searching for particles in nature that might be qualified as being solitons, if only we could identify the field variables and equations that would justify this.

Tsunamis do eventually weaken and disappear, so they are not solitons in a true sense, but one can devise equations that keep their soliton solutions absolutely stable. In solid crystals, one may encounter such situations, for instance if they describe *frustration* in the lattice structure of the crystal. By 'frustration' we mean the following: at large distances away from a region in the center, one may hardly notice that the atoms are attached to one another with a mismatch, but at some points in the crystal the mismatch may stand out. The mismatch itself may look like a particle, but more often it takes the shape of a line, or a surface; in any case, the unnoticeable mismatch far from the center guarantees that the soliton cannot disappear, unless one re-arranges a very large number of atoms, which requires much more energy than what is present in the 'particle'.

A phenomenon in the physical world is said to be topological if one finds some peculiar, stable structure that can only be explained in terms of hardly visible misalignments far away, gently filling all of space. A nice example is *knot theory*: a long piece of rope (a 'one-dimensional world') can look deceptively featureless far away, but if one pulls at its ends, one finds that a structure forms that is locally stable and cannot be undone unless we rearrange the entire rope. This is a knot, and I do not think I need to explain that knots can be complicated to study. Things similar to knots can appear in many branches of physics.

Protons and neutrons are structures in particle physics that are remarkably stable. Indeed, it was noticed that the fields describing pions near these particles carry information about their internal charges (both electric and some other kind of charges called chiral charges). One can devise field equations whose soliton solutions may be identified as protons and neutrons. They are named *Skyrmions*, after their discoverer [11]. Protons and neutrons can also be regarded as being built from up-quarks and down-quarks, and one can understand their stability in other ways. This is typical in physical theories: one often encounters different ways and languages to arrive at the same kind of understanding.

Without the equations, solitons are difficult to understand or even recognise. Thus, when finally the Standard Model of the subatomic particles saw the light, and we understood the equations, more solitons were discovered. A fine example is the *magnetic monopole*. It was first realised by Paul Dirac that, as electric charges always come in multiples of the same fundamental charge that is seen in electrons and protons, one can imagine the existence of pure magnetic charges, but only if they come in multiples of the same quantum, $g_m = 2\pi\hbar/e$, in natural units, where $e$ is the electric charge quantum [12].

Notice that this expression for the unit of magnetic charge contains Plank's constant, $\hbar$. This underlines the fact that the need to have integral units of magnetic charge arises from difficulties in devising wave functions for particles that carry pure magnetic charges; the resolution found by Dirac required a careful analysis of topological properties of quantum wave functions travelling through electric and magnetic field lines.

Dirac did not pursue this idea. Purely magnetically charged particles were never detected, and Dirac could not calculate the mass/energy of these objects. But, when unified field theories for the elementary particles were studied, it was found that the equations could be modified in just such a way that topologically stable solutions would exist. Surprisingly, these solutions carry pure magnetic monopole charges. Their properties, including mass and magnetic charge, could be calculated. Magnetic charge must be absolutely conserved, just as electric charges are, and as the magnetic charge quantum turns out to be large, these particles would really stand out as interesting objects.

However, the adaptations needed in the field equations, have not yet been verified in observations. The terms that would give life to magnetic monopoles would have been a natural further step in the unification of electromagnetism with the weak force. They would at the same time destabilise protons and neutrons. Neither monopoles, nor proton decay, have yet been detected experimentally.

Not only particles may have a topological origin, one can also have topological *events*. For mathematicians, the argument is simple: particles are solitons in three space dimensions, tsunamis and waves in channels are basically one-dimensional. However, depending on your field equations, you can have four dimensional solitons as well. They behave as particles that occur only at one short instant in time, called *instantons*. These were readily identified in the Standard Model. These solitary events can be quite remarkable. For some time in the early days of the Standard Model, there was one particle, called the eta meson, $\eta$, that, according to its equations, ought to behave just like the pions. But it didn't, the eta is much heavier than the pions. The problem instantly disappeared when it was realised that the theory generates instanton events [13]. They can be seen as interaction events exactly of the type that should raise the mass of the eta particle; contradictions with the observations disappeared when this was realised.

Solid state theory is particularly rich in topological phenomena [14]. This is because here, 'space at infinity' is not the vacuum but the fabric of the solid under study, and solids can have many different possible internal structures. But can we attribute *all* features in a solid to topological effects? Of course not, but sometimes phenomena are observed that could have topological origins. In the world of the fundamental particles such questions are particularly intriguing since topology involves properties of the surrounding vacuum itself. Any new piece of insight there can help us understand the world we live in—the vacuum is the same almost everywhere.

It would be fantastic if we could identify more interaction types that may or may not already be familiar in the existing theories, but might be re-interpreted as being topological. Typical for topological interactions is that large amounts of energy, or *action*, to be more precise, are needed to create such knots in space and time. Regarding instanton interactions as tunnelling events, one finds that topological interactions are often extremely weak. Actually, these interactions may be weak in terms of the scale where the topological effect takes place, but they might become sizeable under special circumstances. The mass of the electron might be such an interaction. The electron is the lightest particle that carries electric charge. Its mass could be due to some topological twist, a knot in space and time, just as what we have in magnetic monopoles. One may consider the electron mass in units that should be relevant at the most elementary scale where interactions take place. In terms of those units, the electron is extremely light. *Neutrinos* are lighter still. We do not know where the electron mass or the neutrino mass comes from. It would be sheer speculation to suggest that they are topological, but then, in spite of the beautiful Standard Model, there is still much that we do not understand.

We are often approached by people with beautiful ideas. The problem is then always that what is really needed is a solid starting point from existing knowledge and understanding. This is confirmed by many singular events in the history of science. Wild guesses almost never lead to progress. Deep thinking, without self deception, is the best one can do.

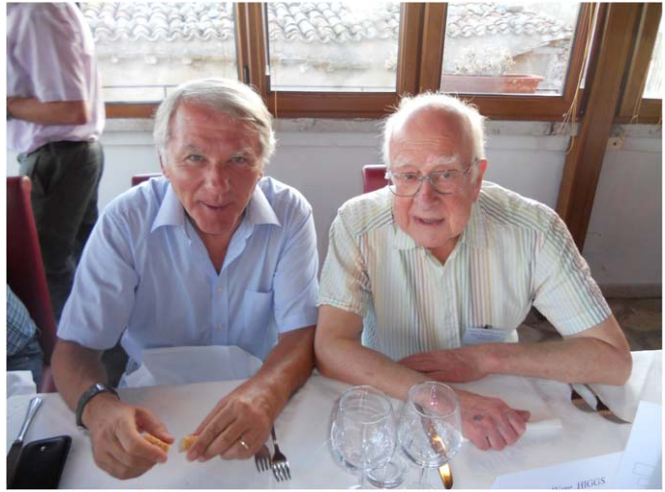

**Figure 3.** Dimitri Nanopoulos and Peter Higgs enjoying the calm before the storm—the following day, Peter Higgs received a phone call informing him that he should attend an official announcement at CERN: The discovery of the Higgs boson was made public on 4th July, 2012.

There is no lack of new ideas or imagination among the newcomers in science. Younger researchers are often inspired to think of new topological issues in all branches of physics. One must realise then that ideas concerning geometric features in the physical world require a solid understanding of the equations we already have, and the models that have been successful in providing understanding of what is going on. The best and most successful ideas usually come from considering the deep and open questions concerning the logical coherence of the theories we have today. There are clashes and paradoxes, but time and again the solutions proposed have been too simple-minded, and did not take all experimental knowledge into account. Needless to stress that the problems we are talking about are hard, just because they still have not yet been solved.

Progress in science seems to slow down just because the unexplored territories seem to be further away than ever. They are still there. Imaginative explorers are welcome to investigate new theories, but only those with the sharpest eyes may stand a chance to show us what still can be done. Eventually, we may discover that geometry and topology are not just words or dreams, they may be the foundations of insights yet to come.

## 4. Will there be new physics? From Classical ->Quantum ->? by Dimitri Nanopoulos

I. We live in very exciting times, 'physics' wise. The discovery [15, 16] of the Higgs boson (see figure 3 where the author is shown with Peter Higgs, after whom the boson is named), the last missing particle of the Standard Model (SM) and the PLANCK satellite data [17, 18] on the Cosmic Background Radiation Anisotropies supporting strongly Inflationary Cosmology, have brought us into a new era of Astroparticle Physics. The opportunities are unlimited, as the combination of LHC experiments and cosmological observations may provide us with 'more than glimpses' towards a Model Of EveryThing (MOET). The theoretical framework that is favored by most of the players in this field is String Theory (ST). While it has not delivered yet, after thirty something years, what a lot of us expected, still for a lot of us, it is the only game in town… Employing Feynman's dictum, 'If you give many reasons in praising a theory, it means that you don't have a great one'; I would only say that String Theory provides a (self-)consistent theory of Quantum Gravity in concert with the other fundamental interactions, strong and electroweak.

Despite this 'rosy' picture, we are facing several rather important and pressing problems, e.g. the Black Hole information loss problem, that bring us directly at the roots of Quantum Theory.

II. Quantum Theory was inevitable in resolving the black body radiation problem, the discrete atomic spectra, … The resolution though was dramatic, because it led us to a completely new physical framework that was not a trivial extension of classical physics. It really changed completely our view of the Universe. If we disregard the historical developments, I believe that the origin of quantum theory is due to the fact that matter is not 'continuous', but is composed from fundamental blocks, that cannot be 'cut' further, the 'atoms'… The Greek word 'atom', introduced by Democritus, was used too soon by Dalton in the 19th century, but one way or another, indicated the existence of fundamental particles in nature.

Having fundamental particles as building blocks, means that we don't have much smaller projectiles to scatter off the fundamental particles and 'see' where these particles are, without disturbing them irreversibly. As such, it is impossible to determine their position, and at the same time their linear momentum, thus making it impossible to define a classical trajectory, as you need the position and the velocity at some time $t_0$! Thus, the idea of probability emerges and the rest is history…

III. The use of the probability amplitude, $\psi$, in the Quantum World leads to the idea of particle-wave duality, and thus the corresponding wave equations (Schrödinger, Klein–Gordon, Dirac…) satisfy the super-position principle, i.e. if $i = 1, 2,…$ n, are solutions of the wave equations, then

$$\psi = \sum_{i=1}^{n} c_i \psi_i$$

$c_i$=complex numbers, is also a solution. Take now a black hole and consider a pair of particles one of which falls into the black hole and the other stays outside. In this case we have no knowledge about the 'fallen' particle and thus we need to sum over all its possible states, thereby essentially turning a 'pure state' ($\psi$) into a 'mixed state' ($\sum_i |c_i|^2 |\psi_i|^2$), absolutely forbidden in 'classical' quantum physics.

That was Steven Hawking's intuitive explanation of the black hole information loss paradox. He proposed that some information is lost and used the idea of a generalized scattering matrix, $, to accommodate this effect. Soon after, I, together with John Ellis, John Hagelin and Marc Srednicki, suggested [19] that we need to abandon the use of $\psi$ (the wave function) and use the density matrix $\rho (\approx \psi \psi^*)$ directly, and we wrote down a generalised Liouville equation for $\rho$ that explained the existence of the Hawking $ matrix. Our starting point, was the idea that quantum gravitational fluctuations, $g_{\mu\nu}$ continuously change the spacetime background metric, thus rendering the use of wave equations impossible and the use of $\rho$ matrices compulsory. With the advent of String Theory, all of the above developments were reconsidered and we have gone through different exuberant and gloomy phases. One day all is solved and understood, the next a problem pops up here and there. I believe that several experts on the subject share my opinion that the jury is still out on the resolution of the black hole information loss problem. The issue being, that yes, if you count all degrees of freedom, Quantum Physics is in full swing, as we learned it as undergraduates, but how is it possible to count all the degrees of freedom if in certain cases we include non-local ones? I argued [20], with John Ellis and Nick Mavromatos, that String Theory contains algebras that support the superposition principle, if everything is taken into account, but effectively this is not possible, and thus we get an 'apparent' loss of information, thereby having the cake (superposition principle) and eating it too ('pure' to 'mixed' state).

I am under the impression that the answer to the fundamental question: *From Classical –>Quantum –>?* will depend very strongly on the type of the resolution that the black hole information loss paradox will have. In other words, if my analysis above about 'effective loss' of information in a black hole environment holds water, then we 'effectively', need to move from $\psi –> \rho (\approx \psi \psi^*)$ as our fundamental entity, thus entering from the Quantum era (S-matrix) to the new quantum era, or in this case the '?' in the question posed in the title above will be replaced by: *Not Quantum —> Classical —> Quantum —> Not Quantum*.

## 5. FAIR—exploring the universe in the laboratory by Karlheinz Langanke

The recent decades have witnessed an exponential growth in our understanding of the world at all scales from the smallest governed by particle physics to the largest spanning the depth of our Universe. From this deeper understanding the exciting insight emerged that both scales are inseparably intertwined as particle and nuclear processes are the drivers of the evolution of the Universe, shaping it from the Big Bang to today and also enabling life to develop on a small planet orbiting an ordinary star. However, every new insight triggers more questions driven by mankind's curiosity and desire to understand the world we live in. Large-scale facilities are one way—sometimes the only one we know—to explore these quests for new science. Here, different strategies are exploited: higher energies (and intensities), improved resolution, better precision. Using the world's most powerful accelerators, CERN has pushed the energy (and intensity) frontiers which culminated in the discovery of the Higgs boson with the LHC [15, 16]. Improved resolution by larger and more sophisticated observatories and instrumentation have allowed astronomers and astrophysicists breathtaking new views of the Universe at all wavelengths, including the detection of gravitational waves [21] and the recent discovery of exoplanets [22]. Improved precision has been behind the spectacular advances made in the laser and quantum optics revolution of recent years (e.g. [23–25]). Precision is also the challenge and the opportunity on the pathway to discover new science in neutrino physics by accelerator-based experiments aiming to determine the neutrino mixing angles, and in this way to explore the matter-antimatter asymmetry in the Universe (e.g. [26]), and in the search for neutrino-less double-beta decay, if observed, proving lepton number violation (e.g. [27]). At FAIR, the international Facility for Antiproton and Ion Research, currently under construction in Darmstadt, Germany, all three strategies will be adopted in the quest for new science and a deeper understanding of the Universe (figure 4). Like in other large-scale facilities, at FAIR new science does not only refer to fundamental new insights, but also to the application of science to new developments serving society.

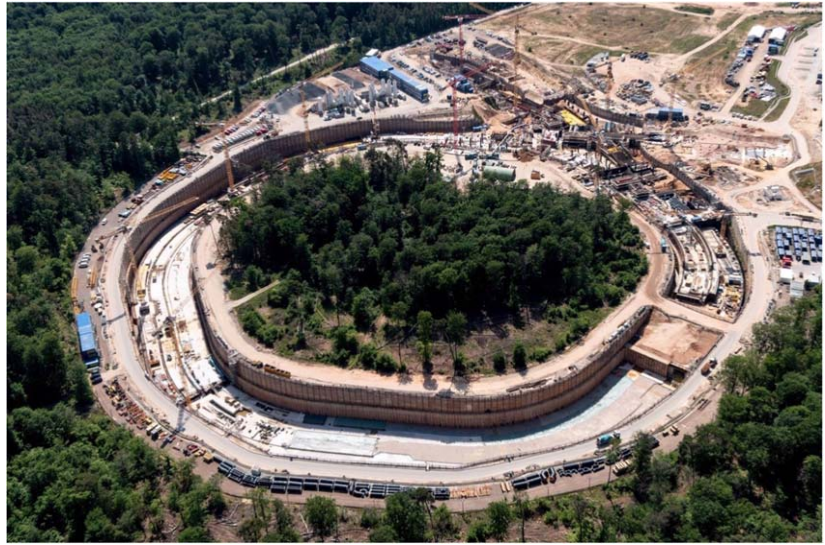

**Figure 4.** The FAIR construction site in spring 2020. The picture shows the progress in civil construction for the SIS 100 tunnel, the 'transfer building', where the beam delivery from the existing and upgraded GSI accelerator chain into the SIS 100 and also the delivery from the SIS 100 to the various FAIR experimental sites will occur, and the cave which will hold the CBM experiment. Now, in winter 2022/3 the SIS100 tunnel is closed and the civil construction on the south campus (upper right part in the figure) has nearly been completed. The first science experiments are scheduled to start in 2027. Credit: GSI.

FAIR is the next-generation accelerator facility for fundamental and applied research providing a worldwide unique variety of ion and antiproton beams [28]. FAIR extends the existing accelerator chain of the GSI Helmholtz Center by a superconducting, fast-ramping heavy-ion synchroton SIS100. This high-intensity machine is supplemented by a proton linear accelerator used for the production of antiprotons, a worldwide unique variety of rings for stored cooled ions (covering more than ten orders of magnitude in energies) and antiprotons, and the Superconducting Fragment Separator for the production and clean identification of secondary beams of short-lived ions. The FAIR accelerator complex is unrivalled by offering beams of all ion species and antiprotons at high energies with unprecedentedly high intensities and quality which are simultaneously available at several experimental areas with a suite of novel detectors and instrumentation for fore-front research in nuclear, hadron, atomic, plasma and nuclear astrophysics, as well as for applications in bio- and radiation physics and material sciences. FAIR is scheduled to start operation in 2027. Until then, the FAIR Phase-0 program, using the upgraded GSI accelerators as well as detectors and instrumentation developed for FAIR, already offers an exciting and unique research program. In the following we will briefly discuss some of the outstanding science opportunities to be exploited at FAIR.

The observation of the neutron-star merger in August 2017 by gravitational waves [29] and by its electromagnetic transient [30] (so-called kilonova [31]) was one of the spectacular scientific highlights in recent years. In particular the kilonova event received a lot of attention as it was the first observational evidence of heavy element production by the r-process related to an astrophysical site. FAIR will contribute to the science of neutron-star mergers and kilonovae in two major ways:

(i) When the two neutron stars merge they create matter of extreme densities (up to three times the nuclear saturation density as observed inside a heavy nucleus like lead) and temperatures (up to $10^{12}$ K, which is about 100 000 times hotter than inside our Sun). At FAIR, such hot and dense matter can be created and studied in ultrarelativistic heavy-ion collisions, as planned by the CBM and HADES collaborations. For the CBM experiment [32], investigating such exotic matter is part of a greater and more general picture: the exploration of the phase diagram of quantum chromodynamics (QCD), the fundamental field theory of strong interaction. Models based on QCD predict nuclear matter to exist in various forms most prominently at high temperatures and/or densities as a new state of matter, the quark-gluon plasma (QGP). We know already from heavy-ion collision studies at the Relativistic Heavy Ion Collider RHIC in Brookhaven and from the ALICE experiment at the LHC/CERN that nuclear matter at high temperatures and zero (net baryon) densities transforms to the QGP phase by a crossover (for a review see [33]). At finite densities, models indicate that the transformation to the QGP should occur by a first-order phase transition [34]. If correct, the nuclear matter phase diagram exhibits a critical point, like water. It is the foremost goal of the CBM experiment to explore the potential phase transition and ultimately to confirm the existence of the critical point. The CBM studies, performed as fixed-target rather than collider experiments unlike at RHIC and LHC or, in the future at NICA, will benefit from unprecedently high event rates achievable with the high energy and intensity beams at FAIR making it possible to explore rare probes and fluctuations as signals for the phase transition.

(ii) The astrophysical r-process produces heavy elements, including the precious metals gold and platinum and all actinides, by a sequence of rapid neutron captures and beta decays (e.g. [35, 36]). The process requires astrophysical environments with extremely high neutron densities, like neutron-star mergers, and involves nuclei with such large neutron excess that most of them have never been produced in the laboratory, including all heavy nuclei relevant for the so-called third r-process peak ('gold-platinum peak') which are essential for the dynamics and the final abundance distribution of the process [37, 38]. This unsatisfactory situation will change in the coming years when the next-generation of radioactive ion-beam facilities will be operational. In particular, at FAIR, with its unique combination of high energies and intensities, r-process nuclei from the gold-platinum peak can be produced and their properties measured for the first time. Thus we will soon witness a gamechanger in r-process nucleosynthesis, placing our understanding on experimental facts, rather than theoretical models.

These activities are, however, embedded in a larger program of the NUSTAR collaboration at FAIR which aims to push our knowledge into as yet unexplored parts of the nuclear landscape with the ultimate goal being to derive a unified picture of the nucleus explaining how the complexity of the large plethora of nuclear phenomena develops from nucleons as the main building blocks and the interaction among them.

In our general understanding, all building blocks of Nature are fermions, while the interaction among them is carried by bosons. For the theory of strong interaction, QCD, these are the quarks which interact by exchange of gluons. As quarks and gluons carry color charge, it is conceivable that in QCD particles exist which are entirely made of gluons (glue balls) or are hybrids of quarks and gluons, which is not possible for other fundamental interactions. QCD also predicts other exotic composite particles like pairs of $q\bar{q}$ or molecules [39]. It is the aim of the PANDA experiment to explore and test these predictions using proton-antiproton annihilation experiments at FAIR [40, 41]. The challenge is, besides producing such exotic particles, to identify their internal structure which, due to models, is reflected in their decay width. This requires, however, the unrivalled resolution feasible with the PANDA detector. Besides opening new doors in hadron spectroscopy, PANDA will also answer specific questions about the internal structure of the nucleon and serve as a factory for hypernuclei, helping to extend the nuclear landscape into the third dimension, spanned by strangeness.

Quantum electrodynamics (QED) is the fundamental field theory of light. Arguably it is the best tested of the fundamental theories, at least in the realm of rather weak fields in which perturbation theory in terms of the expansion parameter $Z\alpha$ holds; here $Z$ is the charge number and $\alpha \approx 1/137$ the Sommerfeld fine-structure constant.

The theory is much less tested for the non-perturbative, strong-field regime as it, for example, applies to the 1s electron in hydrogen-like lead or uranium ions. Stringent tests will be possible in the future using highly-charged ion beams in the FAIR storage rings where precision measurements of the Lamb shift of such exotic ions can be performed. Particular exciting situations of strong-field QED occur in hydrogen-like ions for large charge numbers. If $Z \approx 100$, the electric field strength in the ion exceeds the Schwinger limit which defines the onset of non-linear optics in the vacuum [42]. At even larger charge numbers $Z \approx 173$, the binding energy of the 1s electron in such an ion exceeds twice the electron mass; i.e. an unoccupied 1s electron orbital can be filled after the spontaneous creation of an electron-positron pair leaving the vacuum charged by the remaining positron. An experimental way to create the predicted charged vacuum is by collision of a uranium atom with a uranium ion which is stripped of all electrons [43]. Such tests of strong-field QED are foreseen for the FAIR storage rings by the international SPARC collaboration [44]. Other applications of stored highly-charged ions will focus on the precise determination of the nuclear properties of the low-energy isomeric state in $^{229}$Th, which holds the potential for a nuclear clock with unprecedented accuracy and robustness [45], or the measurement of astrophysically relevant nuclear reaction cross sections. With the FAIR storage rings it will also be possible to realize Heisenberg's idea of a Coulomb explosion [46] in which the electron cloud of a highly-charged and fast moving ion is removed 'instantaneously' by Coulomb interaction with another ion and at extremely low momentum transfer so that the electrons including their quantum-mechanical entanglements can be observed by dedicated detectors.

New science at large-scale facilities can also come as new applications. Arguably the most famous example is the Internet, developed at CERN. At GSI, biophysicists and accelerator scientists combined with physicians from Heidelberg to develop a new accelerator-based cancer treatment—hadron therapy [47]. Originally hadron therapy was applied at GSI to about 400 patients with cancers hardly accessible for surgery. As the 5-year survival rate significantly surpassed those of other methods, two dedicated hadron therapy centers have been constructed in Germany, and one in Shanghai, following GSI's pioneering work. These centers can now routinely treat more than 1000 patients per year. Hadron therapy is an excellent paradigm confirming that large-scale facilities, with their widespread expertise and infrastructures, are particularly suited to develop novel technologies. In the case of hadron therapy this was the joint effort of radiation and biophysicists, accelerator scientists, biologists as well as detector and IT experts. Within the FAIR Phase-0 program, and later at FAIR, several new roads will be explored in accelerator-based treatments of diseases, including heart arrythmia, hadron therapy within the FLASH mode [48], where the curing radiation is delivered within a single high-dose shot, and with further reduced damage to the healthy tissue, and radio immunology (e.g. [41]). Another field in which FAIR, with its unique combination of high energies and intensities, will play a prominent role is connected to space missions, for which the fundamental cross-sections for the interaction of cosmic rays with matter will be determined, in close collaboration with the European Space Agency [41]. ESA has named FAIR its official laboratory for radiation protection studies.

In summary, FAIR brings the 'Universe into the Laboratory' and with its widespread fundamental and applied research opportunities will deepen the understanding of our universe and the objects therein. FAIR will begin operation in an 'along the beamline' approach, with the NuSTAR, CBM and APPA experiments starting first, followed by PANDA after the storage rings have been added to the accelerator complex. In this manuscript, we have briefly summarized some of the expected scientific 'news', the known unknowns, to be discovered and explored at FAIR. And then there are the 'unknown unknowns', in the language of former US secretary Donald Rumsfeld, which are unpredictable and come as a surprise. But they are the greatest fun.

## 6. How does a quantum measurement decide which outcome is observed? by Edward Fry

In the early 1900s, fascinating physical phenomena were discovered that simply did not fit within the broad understanding of classical physics. This led to the development of quantum mechanics; thereby providing an understanding that led to widespread euphoria by the end of the 1920s. Basically, quantum mechanics (QM) predicts probabilities for the specific values that measurements can produce. In the Copenhagen Interpretation, a physical system typically does not have definite properties prior to being measured; but the measurement process affects the system and the result of the measurement is, and has a probability corresponding to, only one of the specific quantum mechanical values that are possible (wavefunction collapse). Although a major contributor to the development of QM, Einstein was one of the few who were concerned and felt that quantum mechanics was incomplete because it could only give probabilities.

As an example, consider a beam of photons traveling along the $x$-axis and incident on a crystal polarizer that is oriented so that vertically ($z$-direction) polarized photons are reflected in the $y$-direction and horizontally ($y$-direction) polarized photons are transmitted and continue along the $x$-axis (i.e. Horizontally polarized photons are transmitted and vertically polarized photons are reflected through 90°). Now, if a photon travelling along the $x$-axis is polarized at 45° to the $z$-axis, quantum mechanics can only tell us that it will be transmitted with 50% probability and reflected along the $y$-axis with 50% probability. Nature somehow makes that decision as to which result will occur, and Einstein felt QM (as it is understood) was incomplete because it did not provide answers to how nature makes the decision. Einstein felt nature should be deterministic, that there must be some additional parameters that would define the result; he did not believe nature could be rolling dice to make such a decision. In fact, in a letter to Max Born dated November 7, 1944, Einstein wrote 'You believe in God playing dice and I in perfect laws in the world of things existing as real objects…' [49, 50].

In 1935, Einstein, Podolsky and Rosen presented an argument that QM was not providing the complete story (referred to as EPR) [51]. Bohm's version [52, 53] of EPR considers two spin one-half particles in a spin singlet (total spin zero) state. The two particles are spatially separated and if the spin of particle 1 is measured in the $z$-direction and found to be in the plus $z$ direction, then one can predict with absolute certainty that measurement of the spin of the spatially separated particle 2 will be opposite (i.e. in the minus $z$-direction). So, the EPR argument is that the spin of particle 2 in the $z$-direction is a real property of particle 2. But if the spin of particle 1 had instead been measured in the $x$-direction and found to be in the plus (minus) $x$-direction, then one can predict with absolute certainty that a measurement of the spin of the spatially separated particle 2 will be opposite, i.e. in the minus (plus) $x$-direction. So, the EPR argument is that the spin of particle 2 in the $x$-direction is also a real property of particle 2. This will be true even if the particle separation is so great that no information traveling at the speed of light could reach particle 2 about the direction of the measurement on particle 1.

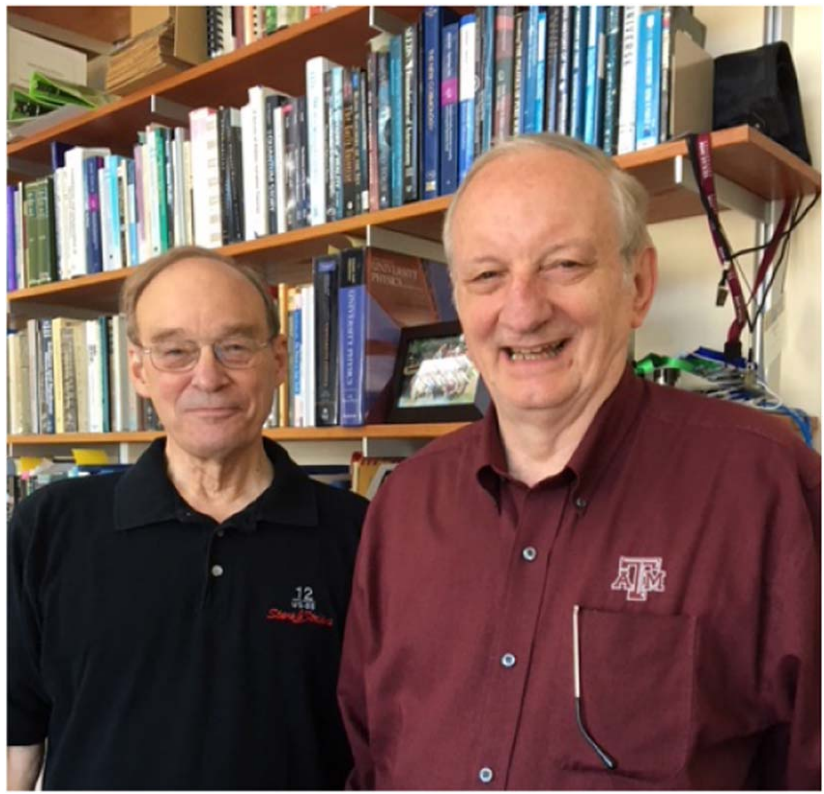

**Figure 5.** Randall Thompson and Edward Fry in recent years. In the early 1970s they were the second group to demonstrate the existence of quantum entanglement via Bell's inequalities. Photograph credit: Ed Fry.

Consequently, the EPR conclusion is that the spin of particle 2 in both the *x*- and *z*-directions is a real property of particle 2. But quantum mechanics does not encompass the existence of real components of the spin of a particle in two different directions; hence quantum mechanics does not encompass all the available physical information; there must be additional parameters that would then enable the replacement of quantum probabilities with deterministic predictions. This is known as Quantum Entanglement, which means the quantum state of each particle cannot be described independently of the quantum state of the other particle. Now, when only measuring the spin of one of the particles, QM only predicts the spin direction with a 50% probability. But when the spin is completely correlated (quantum entanglement) with the spin of another particle, if you measured a spin result for one of the particles, you could make a 100% exact prediction for the measurement result of the second particle spin in the same direction.

All the discussions were philosophical for many years. But in 1964 John Bell showed that any theory that included additional variables and would make deterministic predictions possible would produce statistical predictions that had to satisfy an equality (known as the Bell inequality) [54]. And, he showed that under some experimental conditions, the statistical predictions of quantum mechanics would violate that inequality. So, for the first time, an experimental test was possible. The first, involving polarization correlations between the photons in a Calcium atomic cascade, was completed by Freedman and Clauser in 1972 [55]; it agreed with QM predictions and violated the Bell inequality. This was followed by an experiment at Harvard in 1974 that got the opposite result [56]! But then Fry and Thompson (shown together in figure 5) were able to get funding for their experiment using photons in a mercury atomic cascade; in 1976 they obtained results that agreed with QM predictions and violated the Bell inequality [57]. Their experiment was quite different and interesting in that it used a J = 1–1–0 transition instead of a J = 0–1–0 transition; but most importantly, their experimental design gave a much better signal to noise ratio: they only needed to take data for 80 minutes versus several hundred hours for the previous experiments. At this time, Clauser also repeated a version of the Harvard experiment and obtained the QM result and violation of the Bell Inequality [58]. Many subsequent experiments starting with Aspect, *et al* [59] in 1981 have all agreed with QM. (Aspect's experiments with lasers used the same cascade in Calcium as Clauser's original experiment, but they had an even better signal to noise ratio than Fry and Thompson.) Most recently and for the first time, three experiments have each simultaneously closed the possible loopholes in previous experiments [60–62].

As a result of these experiments, one clearly cannot have some additional parameters ('hidden variables') to get deterministic results for quantum phenomena. In that example of a photon polarized at 45° and incident on a polarizer that transmits vertically polarized light, we have no way of knowing if a specific photon will be transmitted. Even though we know everything presently possible about the photon (e.g. it may be one of the photons from a down conversion pair), we can only say there is a 50% chance it will be transmitted. But Nature does know (or decides) if it should be transmitted; how does Nature decide? Is Einstein wrong? Does God play dice? If so, what is the procedure; what are the dice and how are they thrown? Can we distort the dice to get different results? We have much to learn and it is knowledge that could be expected to have huge impacts on subjects such as quantum information science.

## 7. Is there new physics beyond the standard model? by François Bouchet

Definitively, YES, but what is really the question?

In some sense the question can be taken to mean 'Do we already know all the fundamental laws of the Universe?' When framed like that, the answer is relatively obvious, given the known limitations of what we understand of the Universe; still, it is worth addressing the question a bit more thoroughly to acquaint ourselves with what many contemporary physicists actually do, and hope for.

Let us start by recalling that we gather facts and elaborate *models*, which are collections of hypotheses regarding the constituents of a system under study (e.g. a collection of masses and springs, of wires, capacitors, resistors, or substances, atoms, fluids, gases, or even the content of the Universe itself), their initial arrangement, and their characteristics (i.e. how these constituents behave in response to their environment). The model should then allow the future behavior of the system to be determined. This can be confronted to actual experimental or observational facts. To be successful the model should describe at least some of the facts, with a certain degree of accuracy, and with as minimal a set of these hypotheses as possible.

Progressively, physicists have developed ever more successful models, out of fundamental 'bricks' with known specific characteristics (density, resistance…), and general laws applicable to them, like the laws of mechanics, electromagnetism, or gravity. A model is superseded when a new one is more 'economical', introducing fewer assumptions, and/or describes more facts successfully, e.g. by having a broader range of application. Models are therefore temporary constructs making it possible to interpret known facts and predict new ones. Obviously the more predictive a model is, the better it is! With time, two set of laws with a very broad range have emerged, quantum field theory and general relativity, with each being of particular relevance to behavior on small and large scales.

The so-called Standard Model of particle physics assumes that the fundamental constituents of matter are neatly arranged in types and families (quarks, electrons, neutrinos, photons…) with the specific equations of quantum field theory describing their interactions. In addition (dimensional) numbers must be measured to nail down the specific properties of each constituent and other dimensionless ones to pinpoint the strength of the diverse types of interactions. This model is highly successful, since it makes it possible to describe a myriad of facts with extreme precision out of a restricted set of hypotheses and characteristic numbers.

But we have strong evidence that this model is incomplete. For instance, neutrinos are massless in the Standard Model. But it was found in the '60s that the neutrino flux from the Sun was smaller than would have been expected from the best model of the Sun at the time. This could be explained if the neutrinos were not strictly massless (through a mechanism of oscillations between different neutrino states along their path to earth). This finding was later confirmed by many other experiments.

Another example suggesting incompleteness is given by a property of the characteristics of the particles known as their hypercharge, which are numbers conserved in strong interactions. When the sum of these numbers is taken over all the degrees of freedom of the Standard Model it is found to be zero, and the sum of their cubes is also naught. This strongly suggests the existence of a specific symmetry (technically described by the SO(10) Lie group) whose existence would be very artificial if the world is not described at high energy by a model in which all forces but gravity are unified.

Another way the model of particle physics may be thought to be incomplete is that the theory needs to assume specific values of a rather large set of parameters, both dimensional and dimensionless (e.g. particle masses and interaction strengths) to successfully describe the experimental outcomes. While this set is small in comparison to the very large number of facts described very precisely, one can't help wondering whether these measured parameters could be derived from a smaller list of numbers, in the context maybe of a more fundamental theory which would change our interpretation of 'reality', for instance by replacing particles with small pieces of vibrating strings as the fundamental objects.

The so-called Standard Model of Cosmology has emerged as the other set of hypotheses and laws, and met with incredible success in its own range of application, the cosmos at large scales. Here, again assuming a restricted set of constituents and how they behave under the laws of general relativity, one can reconstitute the evolution of the Universe and understand the formation and evolution of the objects it hosts. It is rather remarkable that such a feat can be accomplished with only a handful of assumptions and the hypothesis that laws derived locally are applicable everywhere, in a realm where they have never been tested before. But again, questions remain, notably as to the nature of the constituents whose existence is inferred from the observations but not (yet?) detected on earth—the so-called dark matter and dark energy—as well as to what happened very early on when the Universe was extremely hot and dense before the ensuing 13.8 billion years of expansion (this is one of the reasons astrophysicists build ever more powerful telescopes).

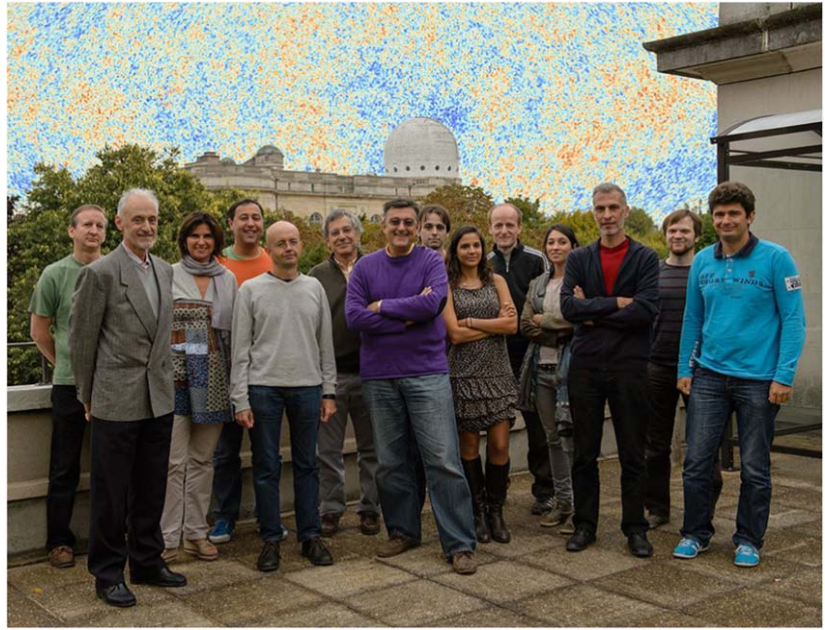

**Figure 6.** In this photo of François Bouchet (in purple) with part of his local team, the sky has been replaced by the fluctuations of the cosmic microwave background (CMB) as rendered by the Planck satellite, popularly known as the echo of the Big Bang. The picture was taken in 2015 on the terrace of the 'Institut d'astrophysique de Paris' where François Bouchet had developed the Data Processing Center which transformed raw Planck data into this map of the CMB. Credit: François Bouchet.

The hypothesis that at very early time, the energy density of the Universe was that of a quantum vacuum which drove a period of very fast expansion, during which irrepressible quantum fluctuations were stretched to cosmological scales is amazingly successful in explaining the origin of the fluctuations which will later condense under the influence of gravity and form galaxies, and lead to their clustering. While this mechanism successfully predicts the properties of the cosmic microwave background anisotropies which surrounds us (and were measured with great precision with the Planck Satellite; figure 6), it calls for an additional component (or several) to the Standard Model. In other words, this Standard Model is an effective model that requires a deeper and more fundamental description of the world.

Addressing these limitations necessitates the development of an improved model encompassing and unifying these two previous Standard Models into a more general one. This will likely require the development of a description of quantum gravity, i.e. of gravity at extreme levels of energy of interaction between the constituents and possibly the introduction of new types of constituents in as yet undiscovered 'dark sectors', or even additional dimensions or spaces inaccessible to most interactions. This may seem a bit outlandish, but who knows what the Universe has in store for us when we probe it as never before?

It is worth remembering that the development of more successful models has historically been achieved by exploring new domains, of energy, duration, etc which unraveled new facts and taught us that there is much more than meets the eye. Indeed, everyday experience provides us with only a very limited view of all the wonderful phenomena that enlarged enquiries bring to our grasp. *Physics is thus really the discovery of the unknown by using our understanding of the already known to develop new technologies and enable further understanding which inevitably leads to new questions, which we then strive to answer.*

So far, the Universe has been very generous with previously unimaginable wonders being discovered every time we enlarged the realm of our investigation, irrespective of whether this was to encompass the extremely small, the extremely large, or even the extremely numerous. Among these incredible phenomena, just think for instance of the relativity of the perception of durations and distances as a function of speed, the quantum phenomenon of intrication or the properties of black holes.

The belief that the Universe is understandable has been met with unbelievable success so far. So why not continue? With this in mind, some physicists strive to develop a better understanding of the world, by exploring new expanses in the hope of developing better models of reality. In other words, we firmly believe that the current Standard Models we have, as successful as they might be, can and most certainly will be superseded by better ones. The only questions are when and how? There is no guarantee that this is just around the corner, or, indeed, that it can happen with the tools we currently use: we might first need to develop much more powerful means before we stumble on key facts that will guide us towards an improved theory.

And we may even ask whether this continued expansion of knowledge is guaranteed in the very long range. Indeed, we recently discovered that the Universe has begun a new phase of progressively accelerating expansion which, if it keeps going on (in the absence of yet undiscovered phenomenon changing the fate of the Universe) will shrink the portion of the Universe from which we can acquire information owing to the constant speed of mediators of interaction… Will this deprive us (if we are still around) from the possibility of discovering some extremely rare new phenomena?

In summary, yes, there is almost certainly new fundamental physics to be discovered beyond what we know, and if history is any guide, we have reason to hope we shall keep unraveling new mysteries shortly, provided we keep looking.

## 8. Will there be new physics? We're not done with the old physics yet by Chad Orzel

The question 'Will there be new physics?' is often interpreted using 'new physics' as a term of art meaning 'fundamental particle physics not captured by the framework of the Standard Model.' There are many reasons, both theoretical and experimental, to expect new developments in this area. On the theoretical side, there is the well-publicized mismatch between the quantum field theory of the Standard Model and the more classical curved spacetime of General Relativity. On the experimental side, the observations of non-zero neutrino masses, the many lines of evidence suggesting the existence of vast amounts of non-baryonic 'dark matter,' and the accelerating expansion of the Universe driven by 'dark energy' all hint at the existence of particles and fields beyond the ones we know already.

In this context, the question is not whether new physics exists—we already have clear evidence that it must exist—but whether we will be able to pin down the exact nature of this new physics in an unambiguous way. This is a tricky question to answer, as there are reasons for both optimism and pessimism. (Many recent books cover aspects of this situation; two that occupy opposite poles are: [63] and [64].) Numerous theoretical approaches have been developed over the last several decades that show promise, and many new experiments in particle astrophysics are coming on-line that may provide experimental confirmation of new particles and fields. It is not clear, however, whether any plausible experiment can definitively pick out any single theory from the vast number of models as the correct model of our universe, a problem that bedevils physicists and philosophers alike.

Returning to the original question, 'Will there be new physics?', though, I would like to construe this more broadly. Fundamental particle physics is without doubt an important subfield of physics, but it is only a subfield. Physics as a whole is a vast subject, spanning a range of scales from the smallest known particles to the size of the entire universe, and there is constant progress and excitement all through that range. Taking a more expansive view makes the answer to 'Will there be new physics?' an unequivocal and enthusiastic 'Yes.' Even if we never discover fundamental particles beyond those in the Standard Model, we will never run out of new discoveries in physics.

Some of the deepest open questions in physics concern not fundamental particles, but the foundations of quantum mechanics: issues of measurement, and interpretation, and the nature of reality [65]. The steady advance of technology is bringing more of these questions within range of experimental tests. 'Cavity optomechanics' techniques coupling the states of a quantum light field with only a few photons to the states of the mirrors confining those photons to a small volume [66] are pushing toward a regime where macroscopic objects can be placed in quantum superpositions. Quantum computer systems that process information with 'qubits' that can be in arbitrary superpositions of '0' and '1' are approaching the number of qubits needed to solve problems beyond the reach of any classical computer [67]. Ultra-cold atom techniques confining atoms within periodic potentials created by light allow quantum simulations of exotic states of matter, with atoms playing the role of electrons in a solid, enabling physicists to study transport properties and phase transitions in unprecedented detail.

Two of the most exciting developments of the recent years (as I write this in May 2019) come from the field of condensed matter physics, and involve exotic forms of superconductivity. One of these involves hydrogen-rich compounds of lanthanum at extraordinarily high pressures—hundreds of GPa—which have superconducting transitions at temperatures approaching the freezing point of water [68]. The other involves paired sheets of graphene rotated by a small angle relative to one another, whose superconducting properties are tunable by varying the rotation angle and the spacing between the sheets [69]. The ability to create new and tunable arrangements of atoms may provide the key to unlocking the mechanism of high-temperature superconductivity in the cuprate compounds, which also feature a layered structure. The origin of the high transition temperatures in these compounds has remained mysterious since the discovery of these materials in the 1980's, so definitive explanation would unquestionably transform our understanding of condensed matter physics, and might serve as the basis for revolutionary new technologies in the future.

Another active and exciting area of research is at the intersection of physics and biology, where techniques developed in physical sciences are driving rapid advances in our understanding of the nature of life. Imaging techniques like cryogenic electron microscopy [70] and super-resolution fluorescence microscopy [71] allow the imaging of biological systems at resolutions down to the single-molecule scale. Even newer developments like lattice light-sheet microscopy can produce nanometer-resolution three-dimensional images rapidly enough to track some biological processes in detail. These provide information about the structure and function of complex biomolecules at an unprecedented level of detail.

Combining this improved understanding of the structure of proteins with information-processing techniques adapted from physical sciences has allowed biophysicists to accurately predict the structure and function of complicated proteins based on their associated DNA sequences [72]; this has dramatic potential both for interpreting genomic data and for developing future medical treatments. Recently developed techniques allow biophysicists to design artificial DNA sequences that self-assemble into arbitrary three-dimensional structures [73]; such tools may enable great leaps in nanotechnology. And at the most fundamental level, statistical mechanics investigations of the entropy of replicating systems may have profound consequences for our understanding of the nature and likelihood of life on Earth and elsewhere in the Universe [74, 75].

In all of these fields, we can reasonably expect that the next 5-10 years will see discoveries with far-reaching consequences for both physics and technology. These expected discoveries are also based entirely on particles and interactions that are already known and well described in the context of the Standard Model. We're not even close to exhausting the potential of 'old physics' yet. So, to close with a return to the original question, whether or not we find new particles and fields, there will unquestionably be new physics in our future.

## 9. What is the dark energy in cosmology? (Hello darkness, my old friend) by Alan Coley and Viraj Sangai

It's been over 100 years since the conception of Einstein's theory of gravity and we are still attempting to fully comprehend it's implications for cosmology. Cosmology is the study of the large scale behaviour of the Universe within a theory of gravity, which is usually taken to be Einstein's theory of General Relativity (GR). GR has been extremely successful in describing observations on small scales, such as the effects of gravity in the solar system. Cosmology is concerned with the dynamics of the Universe on large scales, particularly when small-scale structures, including for example galaxies, are not dynamically significant. Indeed, the Cosmological Principle asserts that on large scales the Universe can be adequately modeled by an exact solution of the equations of GR which is spatially homogeneous and isotropic, which implies that on large enough scales the Universe is assumed to be the same at every point and in every direction in space, respectively (which is clearly not true on the astrophysical scales of galaxies). The standard spatially homogeneous and isotropic Friedmann–Lemaitre–Robertson–Walker (FLRW) model (or the so-called 'ΛCDM cosmology') has been extremely successful in describing current observations. However, it does require the existence of dark matter and dark energy that gravitationally dominate the present Universe but that have never been directly detected observationally.

This implies that if Einstein's theory of GR is truly the best universal theory of gravitation available, then we don't understand what 95% of our Universe is made up of. Of this 95%, about 70% is expected to be dark energy and the rest is dark matter. On the scale of galaxies, gravity appears to be stronger than we can account for using only particles that are able to emit light. So dark matter particles constituting 25% of the mass-energy of the Universe are added. Such particles have never been directly detected. The Universe's dark matter content is approximated using galaxy rotation curve observations, the predictions from nucleosynthesis and computations of the formation of structure. It is not currently known whether dark matter is to be attributed to a particle or describes some modification of GR. However, it is fair to say that it is generally thought that the missing dark matter will be described by normal physics.

Dark energy is motivated by the fact that on large scales the Universe has apparently been accelerating in its expansion for the last few billion years. Gravity, which is a force expected to pull objects closer together, appears weaker than expected in a universe containing only matter. So 'dark energy' is added: a weak anti-gravity force that essentially acts independently of matter. In 1998, the Nobel Prize was awarded for this discovery [76, 77] where supernovae were used to determine the distance to distant objects and, hence, infer the rate of change of expansion of the Universe. Within standard cosmology, the cause of this apparent acceleration is commonly called *dark energy* (with an effective repulsive gravitational force), which has similar properties to a relatively small cosmological constant. Next, we will briefly review the problems associated with determining the nature of dark energy in cosmology.

On first impression, it seems that the most natural candidate for dark energy is a cosmological constant [78]. However, the expected magnitude of a cosmological constant from GR for dark energy energy is incompatible with what is expected from quantum field theory (QFT). This is often referred to as the cosmological constant problem, which is believed to be one of the most fundamental problems in conventional physics [3, 4, 79]. Standard QFT includes an enormous vacuum energy density which, due to the GR equivalence principle, behaves gravitationally in an equivalent way to that of a cosmological constant, which consequently has a considerable effect on spacetime curvature. However, the effective cosmological constant as deduced observationally is exceptionally tiny compared to that consistent with QFT, which implies that a fiducial cosmological constant must balance the enormous vacuum contribution to better than 120 orders of magnitude, for the predictions of QFT to be compatible with GR. It is an extremely difficult fine-tuning problem that gets even worse when the higher-order loop corrections are included, which leads to radiative instabilities. This doesn't just require a one-off fine tuning, but an order-by-order retuning for higher-order loop corrections [80].

In addition, there is the cosmological coincidence problem of explaining why the Universe has started accelerating exactly when it has. This corresponds to explaining why the observed dark energy density is the same order of magnitude as the present mass density of the matter in the Universe, and why dark energy has only just begun to dominate the Universe in our recent history. A proposed solution to this has led to the speculation as to whether dark energy is a pure cosmological constant or whether it is dynamical, perhaps arising from a scalar field model such as quintessence or phantom dark energy [81]. Some physicists have also suggested different reasons for these gravitational effects that do not necessitate new forms of matter [82], but such unpopular alternatives often lead to modified gravity on large scales [83]. It is of interest to ask whether the dark energy problem can be resolved by new physics such as, for example, by including quantum effects, or by old physics such as, for example, classical GR.

### 9.1. New physics approach

Let us now discuss how certain semi-classical and quantum gravity (QG) approaches seek to address the dark energy problem. A homogeneous spacetime with a positive cosmological constant is called a de Sitter spacetime. In a dark energy dominated universe with a cosmological constant, a de Sitter spacetime is required to account for the accelerated expansion. In [84], Friedrich proved a result to show that de Sitter spacetime is a stable solution of Einstein's GR field equations. This is significant for cosmology because it implies that de Sitter spacetime acts as a dynamical attractor for expanding cosmologies with a positive cosmological constant. Also, it is known that [85] any non-collapsing spatially homogeneous model with matter satisfying both the strong and dominant energy conditions will dynamically evolve to an isotropic de Sitter spacetime. Indeed, it can be shown that initially expanding solutions of the field equations of GR with normal matter and a positive cosmological constant exist globally in time [86]. There are also some partial results for inhomogeneous cosmological models with a positive cosmological constant [87]. It should be noted, however, that an accelerated expansion (and, in particular, inflation) in the presence of (an effective) positive cosmological constant is believed to be anti-entropic in the context of Penrose's notion of a gravitational entropy [88, 89]. Gravitational entropy is the concept of applying an analogous form of the second law of thermodynamics to gravitational fields [90].

Recently, in an attempt to understand the compatibility of GR with QFT in the context of cosmology, the stability of quantized de Sitter spacetime with a conformally coupled scalar field together with a vacuum energy has been studied. Indeed, utilizing a semi-classical backreaction it has been demonstrated that a local observer in an expanding Universe does not experience de Sitter spacetime to be stable [91]. Here, backreaction refers to the process wherein a spacetime contains a constant thermal energy density, despite expansionary dilution, due to a continuous flux of energy being radiated from the cosmological horizon, which leads to a late time Hubble rate evolution which differs from that in de Sitter spacetime quite significantly. This seemingly contradicts the thermodynamical treatment in [92] in which, unlike the Schwarzschild black hole spacetime, de Sitter spacetime is argued to be stable. However, if de Sitter spacetime is in fact found to be unstable to quantum corrections, a physical decay mechanism might be possible to significantly reduce the cosmological constant problem (and perhaps also alleviate the fine-tuning in extremely flat, observationally motivated, inflationary potentials).

It is often believed that new physics, from the quantum or classical realm, is needed for a solution to the dark energy problem. However, it is looking increasingly unlikely that a natural solution will be found within QG. Indeed, rather disappointingly, Weinberg and others have adopted the view that, of all of the proposed solutions to this problem, the only acceptable one is the controversial anthropic bound [93]. However, as well as new physics, it is possible that a resolution or at least an alleviation to the problems related to dark energy and dark matter might be sought by studying the effects of small-scale inhomogeneities in cosmology within classical GR more thoroughly, or by some well-motivated modified theory of gravity on large scales. We will now elaborate on this by suggesting how the old physics of understanding Einstein's field equations in an inhomogeneous universe might be crucial to fully understanding the properties of dark energy.

### 9.2. Old physics approach

GR is a local theory of gravity. To obtain the gravitational field equations on large cosmological scales, presumably some form of averaging or coarse graining of Einstein's GR field equations must be performed. Such a spacetime averaging approach must be well posed and generally covariant [94, 95], leading to a well defined way to average tensors in an inhomogenous universe. The averaging of the geometry in GR will consequently lead to an averaged (macroscopic) geometry and enable the macroscopic correlation functions which emerge in the averaging of the non-linear field equations to be computed [96, 97]. There has been some practical progress by using a phenomenological approach of splitting a cosmological spacetime and performing spatial averages over scalar quantities [98, 99]. However, from a mathematical standpoint, a better understanding of the notion of averaging of Einstein's field equations in cosmology is needed.

From an observational perspective, the local Universe is neither isotropic nor spatially homogeneous. Observations indicate a very complicated Universe in which clusters of galaxies of differing sizes constitute the greatest gravitationally bound structures which then form filamentary and two dimensional regions that encompass underdense voids [98]. Indeed, by volume the dominant fraction of the current Universe resides in voids with a characteristic size of about 30 megaparsecs [100, 101]. In addition, any statistical spatial homogeneity of the Universe can only arise on a minimum scale of approximately 100 megaparsecs, and significant variations of the number density of galaxies (on the order of 8%) still still occur in the largest possible surveys [102–104].

In standard cosmology, it is assumed that the background expands as if there are no cosmic structures. Gravitational instability leads to the growth of stars, galaxies and clusters of galaxies, which are simulated computationally using Newton's simplistic theory of gravity. This approach does produce a structure resembling the observed cosmic web in a reasonably convincing way. However, it also necessitates inventing 95% of the energy density of the Universe in the form of dark energy and dark matter to make things work. Even then, the model itself still faces problems that range from tensions to anomalies [105–107], including the existence of structures on gigaparsec scales such as the cold spot in the Cosmic Microwave Background and some super-voids at late-times, and especially the Hubble constant problem [108–110]. These need to be fully understood in the context of the Standard Model of Cosmology, otherwise a non-standard physical explanation is necessary.

It is important to understand the effect of small-scale non-linear structure on the large-scale expansion [111]. After coarse graining a smoothed out macroscopic geometry and macroscopic matter fields are obtained, which are valid on larger scales. Such averaging of local inhomogeneities on small scales can lead to very significant effects on the average evolution of the Universe [98, 99], which is referred to as 'dynamical backreaction'. There is an additional 'kinematical backreaction' arising from the fact that light behaves differently in an inhomogeneous universe in comparison to a spatially homogeneous and isotropic one. For example, inhomogeneities affect curved null geodesics and can significantly alter observed luminosity distances, which are used to infer the accelerated expansion of the Universe [112]. Therefore, averaging (and inhomogeneities in general) can affect the interpretation of cosmological data [113].

While most researchers accept that backreaction effects exist and are important in current precision cosmology, the real debate is about whether this can lead to more than a percent difference from the mass-energy budget of standard cosmology. Any backreaction solution that eliminates dark energy must explain why the law of average expansion appears so uniform despite the inhomogeneity of the cosmic web, something standard cosmology assumes without explanation.To date it is believed that backreaction cannot account for the current (apparent) acceleration of the Universe [114–116]. However, whatever the final resolution of the dark energy problem, it will likely include the important ingredient of classical GR that matter and geometry are coupled dynamically, even at the quantum level [117].

## 10. Are the secrets of the Universe hiding in your bathtub? by Sam Patrick

The possibility of using laboratory-based experiments to simulate quantum fields in curved spacetime was suggested by W G Unruh in 1981 [118] when he demonstrated that the equations for sound in a moving medium are identical to those describing certain fields moving through curved spacetime. Originally proposed as a means of verifying Hawking's prediction of thermal radiation from a black hole, the idea subsequently grew into a new field of research called *analogue gravity* [119], which aims to understand the analogues of various gravitational phenomena in a broad range of condensed matter systems.

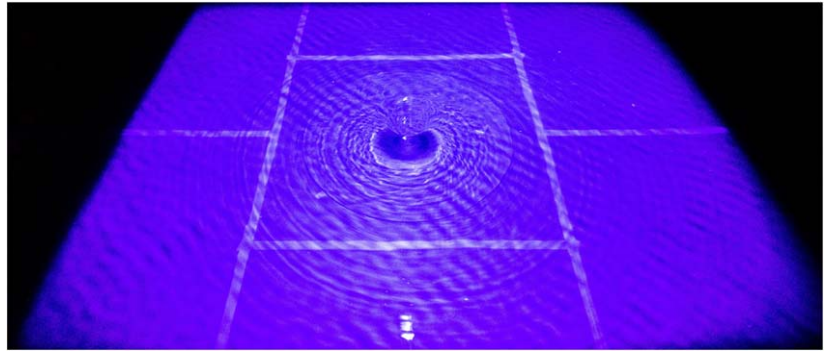

**Figure 7.** Water in a large tank draining through a small outlet in the centre. This phenomenon is commonly called a *bathtub vortex*, and is known to act as an analogue rotating black hole for long waves on the water's surface [120]. Image courtesy of Leonardo Solidoro.

To grasp the concept underpinning analogue gravity, we consider a system with which the reader will (hopefully!) be familiar: water in a bathtub. In particular, think of what happens to waves on the surface of the water draining from your bath after you've pulled the plug (see e.g. figure 7). Since all the water in the tub is being focussed into a small region above the outlet, the flow of water speeds up as it converges on the plug-hole. If the flow is fast enough, there will be a location where the water's speed is equal to the wave speed. Inside of this location, the waves are unable to escape the pull of the drain and instead get dragged down the plug-hole. This mimics the way that light cannot escape a black hole once it crosses the horizon. The draining bathtub is said to be an *analogue black hole*.

Analogies like this arise not only for surface waves in water but in a variety of physical systems, such as sound waves in classical fluids [118], phonons in Bose–Einstein condensates (BECs) [121], light in optical systems [122], ripplons in superfluid helium [123] and polariton fluids in microcavities [124]. Over the past decade and a half, a number of analogue black hole experiments have been created in a diverse range of laboratory set-ups including water flumes [125–127], flowing BECs [128, 129], nonlinear pulses in optical fibres [130] and optical vortex beams [131]. In addition to black holes, another area of high activity is the simulation of phenomena associated with an expanding universe, such as Hubble friction, cosmological redshift and particle production [132, 133]. A detailed historical account of the field is given in [119] and a more recent miniature review can be found in [134].

### 10.1. The Hawking effect

For a long time, the holy grail of analogue gravity was considered to be the measurement of spontaneous Hawking radiation in an experiment; that is, thermal emission from a black hole arising purely out of the quantum vacuum. This stemmed from the fact that Hawking's prediction [135] implied that radiation escaping a black hole would have ultra-short wavelengths near the horizon, and was worrisome since the notion of a spacetime continuum is expected to fail below the Planck length where quantum effects come into play. This put Hawking's prediction on shaky ground and became known as the *Trans-Planckian problem*.

It was in the context of the Trans-Planckian problem that the analogue gravity framework found its very first application [119]. The key idea is that the analogy between fluids and gravity arises at large length-scales where the notion of a continuum fluid flow makes sense. In this regime, the dispersion relation for sufficiently long wavelengths $\lambda$ will be of the form $\omega^2 = c^2 k^2$, where $\omega$ is the frequency, $k = ||\mathbf{k}|| \equiv 2\pi/\lambda$ is the modulus of the wavevector and $c$ is the wave speed (note the equivalence with the relativistic dispersion relation describing electromagnetic waves in a vacuum, where $c$ plays the role of the speed of light). However, fluid mechanics is not a fundamental description of nature since at small enough length-scales, one must account for atomic granularity of the medium. Once the microscopic details of the fluid are taken into account, the dispersion relation receives modifications of the form,

$$\omega^2 = c^2 k^2 (1 \pm \Lambda^2 k^2) + \mathcal{O}(k^6), \tag{1}$$

where $\Lambda$ is a small length-scale marking the onset of the microscopic physics. The key insight, provided by Jacobson, is that this mimics our expectation in gravity that new physics should arise below the Planck scale [136].

The consequence of the modified dispersion is that the group velocity $\partial\omega/\partial k$ is no longer a constant $c$ but now becomes $k$ dependent. When one takes the $+$ sign in equation (1), short wavelengths travel faster than $c$ and the radiation emerging from an analogue black hole originates inside the horizon. With the $-$ sign, short wavelengths are slower than $c$ and the radiation starts as an in-going wave outside the horizon. Early numerical simulations (and subsequent analytic studies) employed such modified dispersion relations to show that the radiation escaping an analogue black hole is remarkably close to thermal for frequencies and temperatures which are smaller than the relevant scale set by $\Lambda$ [137, 138]. This gives one confidence that the Hawking effect should still occur for real black holes, in spite of our ignorance of physics below the Planck length.

The next natural question was whether the thermal spectrum from an analogue black hole could be measured in the laboratory. To this end, a series of experiments were performed using a BEC accelerating through a waterfall type configuration (see [129] and references therein). These results remain somewhat controversial, as it has not yet been agreed whether the radiation occurs spontaneously or whether other classical noise sources are at play [139]. Nonetheless, the fact that features of the Hawking effect can arise in such a manner has been argued to be a remarkable and unexpected discovery unto itself. Related to this are classical experiments involving surface waves in open-channel flows where the stimulated Hawking effect can arise due to coherent input signals [125, 126] as well as turbulent noise on the water's surface [127].

**10.2. Beyond the Hawking effect**

In recent years, the scope of analogue gravity has broadened significantly. The aims at present are numerous owing to the diversity of systems encompassed by the field. Roughly speaking, however, these goals fall into one of two categories: those which are system-oriented and those which are gravity-oriented.

*10.2.1. System-oriented studies*

System-oriented studies aim to learn something about the physics specific to the analogue system being used. There are several reasons why this is interesting. I will illustrate some of these in the context of the experiments of [120] where an effect called *superradiance* was measured using the bathtub apparatus depicted in figure 7.

Superradiance is a close relative of the Hawking effect, involving the extraction of energy from rotating or charged systems through the amplification of incident radiation. The basic mechanism behind superradiance can be understood using a simplified model of the bathtub vortex in which the water flows with velocity $\mathbf{v} = -D/r\,\vec{e}_r + C/r\,\vec{e}_\theta$. This says that the flow speeds up as the distance $r$ from the plug hole decreases, where $C$ and $D$ are positive constants that determine the circulation and drain rate respectively. Ripples on the water's surface with long wavelengths propagate at an approximately constant speed $c = \sqrt{gh}$, where $g$ is the acceleration due to gravity, $h$ is the water's depth and a 'long' wave in this context means $kh \ll 1$. The region where $\|\mathbf{v}\| > c$ is special since here, waves with positive frequency $\omega$ can have a negative frequency in a reference frame co-moving with the fluid, i.e. $\Omega = \omega - \mathbf{v}\cdot\mathbf{k} < 0$ is allowed. Since the sign of the wave energy is related to the sign of $\Omega$, this means that a wave which started with positive energy far away from the vortex can have negative energy as it goes down the plug-hole [140]. But because energy in the system is conserved, this implies there must be a reflected wave escaping from the vortex which has more energy than the wave that was sent in. In other words, the wave gets amplified, and the energy required for this amplification is extracted from the system when it absorbs the negative energy.

Perhaps the most important lesson to come from the detection of this effect in the laboratory was how resilient it is against non-idealised conditions. In particular, the water wave analogy with black holes is only mathematically precise for shallow water waves (i.e. those with $kh \ll 1$) in an inviscid, irrotational fluid [141]. In any real experiment, all of these assumptions will be broken to varying extents. Most surprisingly, the detection of superradiance in [120] was actually performed for deep water waves, which satisfy $kh \gg 1$. These are strongly dispersive waves with the approximate dispersion relation $\omega^2 = g|k|$, which is not of the perturbative form of equation (1). Hence, it is quite remarkable that a phenomenon anticipated within the regime of the analogy should still occur so far outside its domain of validity. It has since been demonstrated that a modified form of superradiance occurs for deep water waves [140], although the influence of having a rotational and viscous fluid remains to be fully understood. This is a common theme in many analogue gravity studies, where differing system properties lead to different incarnations of the effects under scrutiny. The gravitational analogy (based on a simplification of the system) motivates experiments, which in turn highlight aspects of the full fluid dynamics that are poorly understood. This then leads to theoretical investigations, opening the door to new physics.

Another theme in the analogues is that progress in one system can spur on developments in others. Following the measurements of [120], the search continues for signatures of superradiance in non-linear optics [131], vortices in superfluids [142] and sound scattered by rotating disks [143]. Understanding how this phenomenon occurs in finite-sized systems is important since the trapping of superradiantly amplified modes can lead to instabilities, which has consequences for the complete non-linear evolution of a system. For example, it has been argued that vortex fragmentation in superfluids [142] and polygon instabilities around classical vortices [144] result from the trapping of superradiant modes. This illustrates how the analogy can lead not only to new discoveries, but also to new interpretations of recognised phenomena. All in all, it is a general misconception that to find new and exotic physics, one has to peer out into the depths of the cosmos. Next time you hop out of the bath, just think that some of this physics might be happening (literally) right under your nose!

*10.2.2. Gravity-oriented studies*
In the second class of studies, analogue systems are used as a test bed to extract lessons for real gravity. This is done principally to learn about the quantum mechanical behaviour of gravity (or the gravitational behaviour of quantum mechanics depending on who you ask!), in view of the absence of a theory which marries general relativity with quantum field theory. These studies are faced with a sizeable problem right from the get-go: analogue gravity is a framework which equates the dynamical equations describing waves moving through curved spacetimes and fluid-like media. However, it does not (in general) equate the dynamical behaviour of the spacetime itself to that of the fluid. There are, nonetheless, still lessons to be gleaned from this line of enquiry.

One can approach this problem by studying the *backreaction*: namely, the influence of fluctuations on the background they propagate through. All non-linear systems exhibit an intrinsic backreaction. For example, in the bathtub set-up in figure 7, waves push water down the plug hole and reduce the total volume of fluid in the system, thereby changing the effective spacetime perceived by the waves [145]. Backreaction studies are particularly interesting in quantum systems, since they have the potential to reveal how fluctuations interact with quantum degrees of freedom in the underlying geometry. BEC analogues have received the most attention in the literature due to their simplicity and inherent quantum behaviour. For example, calculations employing BECs have been used to show:

- Backreaction approximations frequently employed in semi-classical quantum gravity do not always give the correct result [146],

- The black hole information paradox (see e.g. [147] for an overview) can be addressed in an analogue system due to the entanglement of Hawking radiation with the mean-field condensate that gives rise to the effective spacetime [148],

- Analogue gravitational dynamics can emerge from the microscopic theory describing the condensate (in the same vein that fluid mechanics emerges from interactions of $10^{24}$ atoms) [149],

- Quantum superpositions of analogue spacetimes are highly unstable, which suggests why we do not observe them in nature [150].

In summary, analogue gravity is by no means a recipe to solve long-standing problems in quantum gravity. But it often happens in searches for new physics that if we aren't getting any answers, we aren't asking the right questions. And this kind of analogous thinking is very good at prompting us to carefully consider what questions we're asking.

## 11. Will there be new physics? by Jim Baggott

Will there be new physics? Most certainly. Despite what some doomsayers might have once wanted to argue, we are not yet at the end [10].

But this is not quite the question, is it? Though it might seem simple and really rather straightforward, this is a question that needs some unpacking. For one thing, it's directed at new 'foundational' physics, of the kind that transcends the current Standard Models of particle physics (founded on quantum field theory) and inflationary Big Bang cosmology (general relativity). In disciplines such as solid-state physics and quantum information, new physics is happening all the time.

Whilst I anticipate that there will indeed be new foundational physics, I can't tell you if new discoveries will be made during your lifetime, or whether these will in any way resemble the speculations of contemporary theoretical physicists. This might seem an oddly ambiguous conclusion given the recent successful discoveries of the Higgs boson and gravitational waves. Until we realise that these discoveries are all *supportive* of the current paradigms: they do not (yet) help us to transcend them. And future (rather expensive) experiments currently at the evaluation, planning, or commissioning stages travel more in hope than in expectation of new foundational physics.

Why is this? Here, I think, there is a simple answer. *Contemporary foundational theoretical physics is largely broken* [63, 151–154]. It offers nothing in which experimentalists can invest any real confidence. Theorists have instead retreated into their own fantasy, increasingly unconcerned with the business of developing theories that connect meaningfully with empirical reality.

About forty years ago particle theorists embarked on a promising journey in search of a fundamental description of matter based on the notion of 'strings'. Lacking any kind of guidance from empirical facts, forty years later string theory and the M-theory conjecture are hopelessly mired in metaphysics, a direct consequence of over-interpreting a mathematics that looks increasingly likely to have nothing whatsoever to do with physical reality. The theory has given us supersymmetric particles that can't been found [155]. It has given us hidden dimensions [156, 157] that may be compactified at least $10^{500}$ different ways to yield a universe a bit like our own [158]. And at least for some theoretical physicists who I believe really should know better, it has given us a multiverse – a landscape (or swampland?) of possibilities from which we self-select our universe by virtue of our existence [159, 160].

Cosmic inflation was introduced as an elegant fix for the flatness, horizon, and monopole problems but in truth it simply pushed these problems further back, to the initial conditions of the Universe at its Big Bang origin. Instead of fretting about the fact that these initial conditions are likely to remain forever elusive, at least within the context of the Big Bang model, why not simply render them unimportant or irrelevant? Why not assume eternal inflation, giving us a multiverse with an infinity of different sets of initial conditions [161–163].

Although the history of theoretical physics reveals a general tendency towards such higher speculations [164], I'm pretty sure there was a time in which this kind of metaphysical nonsense would have been rejected out-of-hand, with theorists acknowledging the large neon sign flashing WRONG WAY. There was surely a time when theorists would have been more respectful of Einstein's exhortation: 'Time and again the passion for understanding has led to the illusion that man is able to comprehend the objective world rationally by pure thought without any empirical foundations in short, by metaphysics' [165]. Alas, instead we get a strong sense of the extent to which foundational theoretical physics is broken. Both string theory and eternal inflation fix on a multiverse and the anthropic principle as 'the solution'. This is judged by far too many influential theorists working at some of the world's most prestigious institutions as a virtue, rather than a vice [6, 166–168].

I believe real damage is being done. At a time when new ideas are desperately needed, the dominance of one particular research programme (no matter how fragmented) is extremely unhealthy. Other approaches, if not to a theory of everything then at least to a quantum theory of gravity, are dismissed or treated as poor second cousins, with the unwavering mantra that string theory is 'the only game in town' [169]. Perhaps conscious of the fact that these parts of contemporary theoretical physics no longer show any interest in empirical data, some theorists prefer to reinterpret the scientific method on their own terms, based on notions of 'non-empirical theory confirmation' [170].

In the meantime, popular science periodicals feature an endless stream of multiverse stories, pandering to an audience that may no longer be able to differentiate science from fringe science or pseudo-science. The very credibility of science is under threat, at a time when public trust in science and scientists is needed more than ever [171].

Yes, there will be new physics. Just don't expect current developments in foundational theoretical physics to offer any clues anytime soon.

It is then legitimate to ask: 'If you're so sure there will be new physics, and it's not going to come from the theorists, where *will* it come from?' Lacking any kind of crystal ball, we are left to speculate. Historical progress in some scientific disciplines can sometimes look like climbing a rope, hand-over-hand. There are moments in history when the left hand of empirical data reaches up along the rope, pulling science upward and leaving the theorists to play catch up. And there are moments when the right hand of theory reaches higher, encouraging the experimentalists to determine if the theory works, or not. The history of twentieth-century cosmology provides a nice illustration of this rope climbing act [172]. In this article I've argued that for the last few decades the right hand has been flailing around, unable to get a purchase on the rope and so unable to pull science in the right direction.

We therefore need to look to the left hand—to experiment—to pull us up out of this impasse. As to precisely where to look, my instinct is to avoid quantum mechanics. It is almost 100 years since Niels Bohr delivered his lecture on the shores of Lake Como, in which he befuddled his audience with his description of 'complementarity'. Nearly 100 years later we're still debating the status of the quantum wavefunction and, to my knowledge, there are simply *no* experimental data judged to be at odds with the predictions of the theory.

The same is not true of inflationary Big Bang cosmology, with its mysterious dark matter and dark energy, which together account for a mere 95% of the mass-energy of our universe. There exists the real possibility of disagreement, or at least a *tension*, between 'early universe' *predictions* of the Hubble constant derived from model-dependent analyses of temperature fluctuations in the cosmic background radiation, and 'late universe' *measurements* of the Hubble constant derived from observations of Cepheid variables and Type Ia supernovae in distant galaxies. If it exists, the tension is small (about 7%–8%). Adam Riess has compared the situation to a civil engineering project that has gone disastrously wrong. Imagine the construction of a (metaphorical) bridge spanning the age of the Universe, begun simultaneously on both 'early' and 'late' sides of the divide. Foundations, piers, and bridge supports have been completed, but the engineers have now discovered that the two sides do not quite meet in the middle [173].

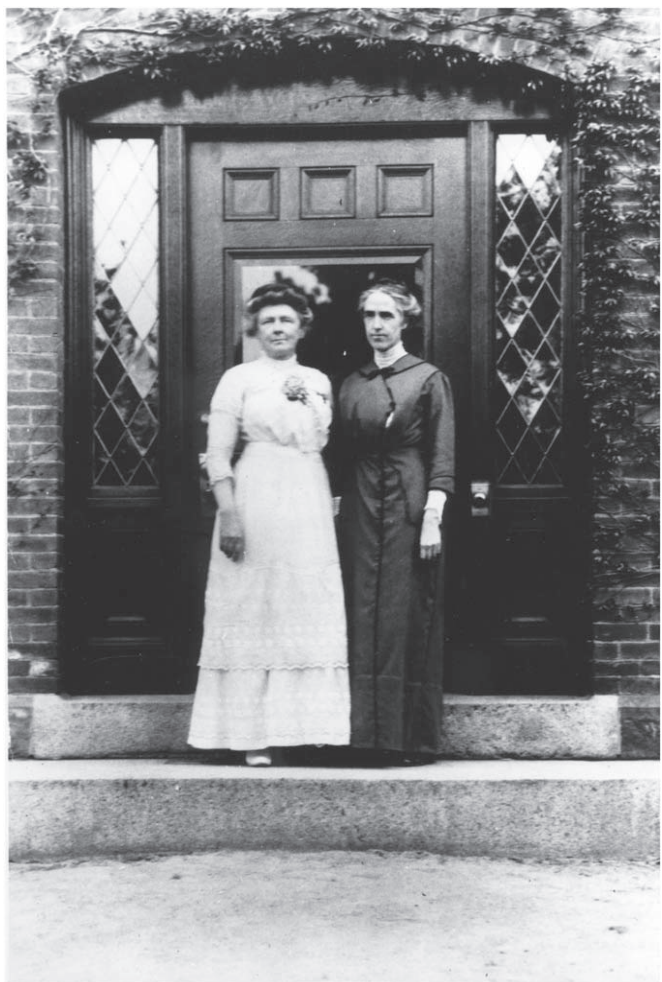

**Figure 8.** Annie Jump Cannon with Henrietta Swan Leavitt, 1913. Their work provided a foundation for much of the 20th century astronomy which vastly expanded our view of Nature. Cannon manually classified a total of around 350,000 stars, and her stellar classification system, adopted by the the International Astronomical Union in 1922, is still being used. Henrietta Leavitt studied the images of 1,777 variable stars and discovered that the time period over which the brightness of the star varies is an accurate measure of the star's intrinsic brightness. This critically important discovery led to many other major discoveries in astronomy by Edwin Hubble and others, including the fact that the Universe is expanding and that are galaxies outside the Milky Way. Credit: AIP Emilio Segre Visual Archives, Shapley Collection.

The evidence is qualified, and not all astronomers agree the extent of the tension, but instruments aboard the James Webb Space Telescope should soon provide clarifying answers (it is already revealing fully-formed galaxies that theoretically should not exist). Naturally, the theorists have already been at work on a variety of ways to bridge the gap [174] . Potential solutions such as Early Dark Energy would seem to compound existing mysteries rather than provide an explanation. But whenever there is disagreement between theory and data, there is at least the prospect (if not the promise) of progress, eventually. And, in these circumstances, it is difficult to imagine how progress will be possible without some form of new physics.

## 12. How big is nature, and how much of it can we explore? by Roland Allen

In the 'Great Debate' of only a century ago—on April 26, 1920—astronomer Harlow Shapley argued that the Universe (as he defined it) consists entirely of the Milky Way [175]. Now, as a result of heroic efforts by other astronomers like those in figure 8, and later Shapley himself, it is known that we live in an enormous universe that spans hundreds of billions of galaxies. This is a recent example of how we tend to underestimate the scale of Nature, while overestimating our own importance and centrality.

On the other hand, we also have to be wary of another fallacious human tendency, which has been equally prevalent throughout history—the inclination to invent extravagant fantasies which are satisfying but unrelated to reality.

Here, as we consider the recent surge of interest in various candidates for a multiverse, let us attempt to evade the Charybdis of mindless reactionary opposition and the Scylla of self-indulgent fantasizing. It is useful to adopt the classification scheme of Max Tegmark [168, 176], who has given due credit to the principal originators. But it is also helpful to break some of his levels into different versions, from most to least convincing. Each multiverse at a given level contains the others at lower levels.

With the definitions given below, we will argue that:

 (i) A person who is fully knowledgeable about the subjects is *compelled* by logic to accept the reality of multiverses 1- and 3. Furthermore, not to accept these views of Nature may be potentially harmful to the progress of science, in the same way that not accepting evolutionary biology would be potentially harmful to biology and medicine, and not accepting the Copernican interpretation of planetary motion would have harmed the progress of astronomy.

 (ii) A fully informed person will also find multiverses 1 and 2 quite plausible—but reservations are understandable.

 (iii) Multiverses 1+ and 4 are worthy of consideration but far removed from present-day science.

There is insufficient space here to do justice to the hundreds of important papers on this subject, but the multiverse concept has received such wide attention (especially during the past decade) that the main references are easily found on the internet, along with reviews, news articles, and videos, and the leading work prior to 2014 is credited in [168].

**12.1. The level 1 multiverse exists beyond our horizon**

A level 1- multiverse—with an expanse reaching far outside our observable universe—has now become just as compelling as a spherical Earth. i.e. after the remarkable astronomical discoveries of the past 25 years, arguing against a level 1- multiverse would be just as plausible as arguing for a flat Earth with edges a few centuries ago.

The observed flatness of our observable universe, plus the observed acceleration of its expansion, implies a vast region beyond our event horizon that we will never be able to observe directly. The full extent of space then deserves to be called a multiverse, inhabited by many other parallel universes having the same laws of physics as our own, but very different outcomes in their cosmological structures and historical development.

When the quite credible theory of inflation is added, this region is further expanded by many orders of magnitude. Whereas a 1- multiverse might contain more than a million universes like our own, a level 1 multiverse might contain more than $10^{75}$, since inflation requires expansion by a factor of about $10^{25}$ or more.

Furthermore, it is not required that space have positive curvature. If it has zero or negative curvature, space will be infinite. Tegmark has noted that an *exactly* flat universe would imply infinite extent with an infinite number of parallel universes. He has further noted that the laws of physics seem to imply *ergodicity*, so that all possibilities would be almost precisely realized—or even precisely realized—or even precisely realized an infinite number of times—if one takes into account quantum limitations on possible states. We regard such a 1+ multiverse as being a fascinating but less than fully convincing hypothesis.

As Tegmark and others have pointed out, even if we cannot *observe* a parallel universe, we can *infer* its existence if it inevitably follows from a theory and set of observations that have achieved the status of being completely trustworthy (in any reasonable sense). The theories and techniques used in cosmology have collectively reached this status, in the present context, even if separate components are still being challenged. Most informed members of the astronomy community would agree that there is a type 1- multiverse as defined here (although many might prefer to use different terms). And it appears that a majority of the community would accept inflation, with its implication of a truly vast number of parts like our own observable universe.

Is it possible for a sufficiently advanced civilization to communicate or travel between different parts of a level 1−, 1, or 1+ multiverse? Since they lie on the same spacetime manifold, and since this manifold can in principle be connected by Einstein-Rosen wormholes, communication or transport is in principle achievable if a technology is developed to create and traverse wormholes. According to the work of Thorne and coworkers and others [177], this must include an exotic antigravity mechanism to prevent collapse of the wormhole.

However, even if such a technology could be achieved, there is another limitation similar to the one implied by the results of Morris, Thorne, and Yurtsever [178]. Namely, a wormhole of the kind they consider must have been created, with its ends placed separately in the two universes, before they are causally disconnected. For two universes which are currently well separated in an inflation scenario, this means within about the first $10^{-32}$ second of the Universe's existence. Since advanced civilizations cannot evolve in $10^{-32}$ second the only possibility for traveling across to another place in the type 1 multiverse would be discovery of a primordial wormhole that was somehow created by exotic processes in the early universe, and then preserved by exotic physics.

Can our remote descendants nevertheless overcome these limitations and develop the technology to travel across the level 1−, 1, or 1+ multiverse? This would require somehow shooting a wormhole across from our place on the spacetime manifold to a place perhaps $10^{12}$ or $10^{20}$ light years away, and puncturing the manifold there to provide an entry for the wormhole. This would require physics far beyond anything currently being discussed in serious publications. But physics has come so far in the past two centuries that it is impossible to put firm limits on what might be achieved in the truly distant future.

If one could in fact project wormholes across the spacetime manifold, there would, of course, be more mundane applications like time travel and rapid transport across galaxies.

**12.2. The level 2 multiverse explains why we can exist**

Suppose that there is a 'primal theory' underlying current physics, in which the laws (including physical constants) are ultimately determined by the structure of an internal space of some kind. A particular version of this structure essentially acts as the genome for a universe, in which it is embedded at every point (just as the genome of a human being is embedded in each cell). The complete set of possible internal structures yields an ensemble of universes, and this is a type 2 multiverse. String theory can be regarded as a toy model of such a primal theory, with a 6-dimensional internal space and a landscape of $10^N$ internal spaces and universes, perhaps with N ∼500.

A primal theory with a large multiverse is plausible because it can explain why so many things in our universe are just right for our existence, in the same way that our planet Earth is just right for life to evolve and survive—unlike nearly all of the thousands of others that have been discovered—and even in the same way that particular regions on Earth can sustain abundant life.

The type 2 multiverse has actually produced one successful prediction, for the approximate density of the dark energy [93, 179, 180].

How can we possibly envision exploring another universe with different laws? There is an obvious fundamental principle:

- One part of a multiverse can be accessed from another part only if they can somehow be connected.

This appears to mean that a different part of a level 2 multiverse could only be reached by a probe which somehow passes through the internal space, on a length scale that is presumably comparable to a Planck length. An ordinary probe *into* the internal space would be a dead-end trip, like a mission into a black hole. What is required is a journey *through* internal space and across to another place in the multiverse—through a topological funnel in a $D$-dimensional manifold that is analogous to a wormhole in 4-dimensional spacetime, except that the external spaces on the two sides can have different numbers of dimensions and different laws. Such a topological object might be called a 'rabbit hole' because it would lead to a such an alien world. Creating or finding such an object, and making use of it, is a task for a supremely advanced technology.

**12.3. The level 3 multiverse is required by quantum mechanics**

A half century ago, when the present author published a brief positive comment on the Everett interpretation of quantum mechanics [181], this interpretation was dismissed and even ridiculed by nearly everyone, as too bizarre to take seriously. The Copenhagen interpretation was generally accepted and regarded as noncontroversial—perhaps because Niels Bohr, shown in figure 9, was so widely revered, or perhaps because of its nebulosity. Hugh Everett, although now admitted by many to the pantheon of genius, was originally so unappreciated that, after a relatively early death, his ashes were discarded in the trash (by his wife, in accordance with his own wishes). This may exemplify the effect of personality on historic developments (figure 9).

More recently, in an informal poll by Tegmark at a 2010 Harvard talk. the outcome was 0 for the Copenhagen interpretation, 3 spread over a set of other heavily promoted interpretations, 16 undecided or 'other', and 16 for the Everett interpretation [168]—indicating that Copenhagen and Everett have swapped positions among those who have thought carefully about this issue.

It was never clear what the Copenhagen interpretation really meant, but a common interpretation of this interpretation is the more specific ensemble interpretation [182]: A quantum state describes only an ensemble of similarly prepared systems, and not an individual system. According to this assertion, there is no place in physics for individual systems. It is amazing that this could have been accepted as the standard view, when it is in flagrant logical contradiction with the actual language and practice of physicists—who commonly refer to the quantum states of individual systems, from particles and ions in traps to macroscopic materials.

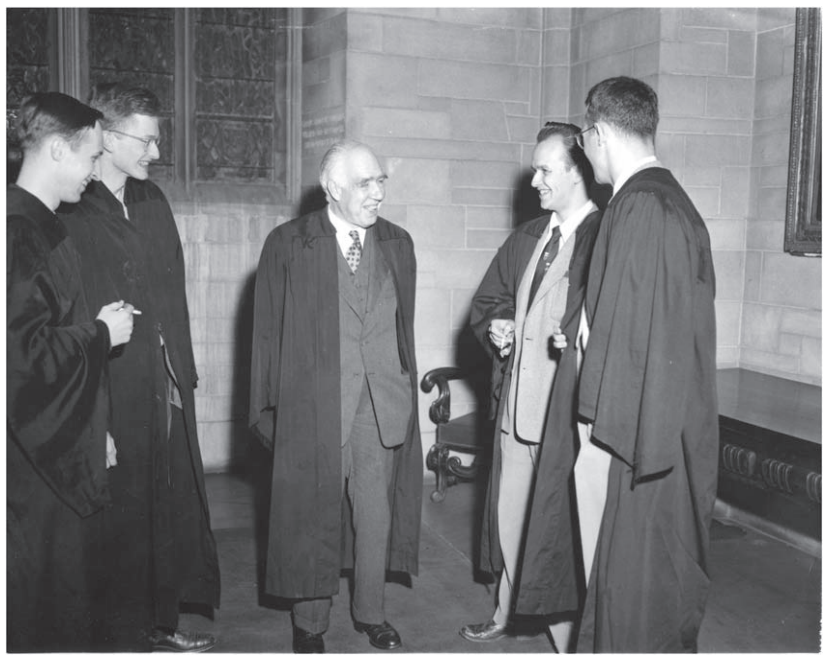

**Figure 9.** Hugh Everett, author of the Everett interpretation of quantum mechanics, standing immediately to the right of Niels Bohr, presumed author of the Copenhagen interpretation. ('Right' and 'left' here are as seen by a viewer of the photo.) On the far left is the distinguished relativist Charles Misner, and to his right the distinguished mathematician Hale Trotter, with David K Harrison at the far right. Photograph by Alan Richards, courtesy AIP Emilio Segre Visual Archives.

Einstein was not so much disturbed by the *indeterminism* of quantum mechanics as he was by its *incompleteness*, as defined by the following statement: '… every element of the physical reality must have a counterpart in the physical theory. We shall call this the condition of completeness.' Einstein saw that quantum mechanics in the Copenhagen interpretation fails to satisfy this reasonable condition. We might add that it also leads to *inconsistency* in how we describe the world using the language of physics.

Another prevailing view was that an individual system has a quantum state, but that this state somehow collapses to a single outcome during a measurement process—and it is again amazing that such a patently absurd idea could have been accepted for so long by so many people.

The Everett interpretation states that quantum mechanics can be accepted just as it is, with no need for embellishments like wavefunction collapse or philosophizing [183]. But the attempts to evade it have produced a torrent of verbiage over the past half century, in a large number of articles, books, and discussions having no scientific impact. The title of a talk and paper by Tegmark neatly summarizes the choice between a clean, well-defined interpretation and the many nebulous attempts at alternative descriptions: 'Many Worlds or Many Words?' [184].

Objections to the revolution in thinking required by the level 3 multiverse are reminiscent of objections to the Copernican and Darwinian revolutions. In each case, there have been byzantine intellectual constructions to avoid a picture that is far simpler, but counter to misplaced intuition, in trying to understand the motion of the planets, the fossil record, or, in the present case, wave-particle duality.

How can we explore other parts of the level 3 multiverse? It is hard to imagine how we can overcome the decoherence of our macroscopic worlds and probe a different Everett branch. Even if we had a wormhole that extended back to a past time $t$, we could only start a new Everett branch at $t$ rather than joining the previously existing branch. But we can fantasize that a technology of the very remote future, perhaps with a nonlinear or other exotic extension of quantum mechanics, might be able to tunnel across Hilbert space to a different state vector.

### 12.4. The level 4 multiverse consists of all that is possible

Multiverses 1, 2, and 3 all live on the same basic spacetime manifold as ourselves (extended to $D$ dimensions). But a mathematical physicist can imagine universes based on other mathematical structures, and even that all mathematical possibilities are realized. According to Tegmark's formulation [168], 'The Mathematical Universe

Hypothesis implies that mathematical existence implies physical existence. This means that all structures that exist mathematically exist physically as well, forming the Level IV multiverse.'

At this point, of course, one has passed far beyond the requirements of observation and logic, into a mode of thinking that is natural for a mathematical physicist, but to others will suggest that the roles of mathematics and the reality of Nature have been reversed—since it appears that mathematics, a human construction founded on human experience, is embedded within Nature, rather than the other way around.

The current status of the level 4 multiverse, therefore, is that it is can be a source of inspiration (and entertainment), but is far removed from normal science. Nevertheless, we can reflect for a moment on what it would mean to explore another type 4 universe—for example, one based on cellular automata [185, 186]. According to the fundamental principle above, we can reach another part of the level 4 multiverse if it can somehow be connected to our universe. Since we are now thinking within a purely mathematical context, this means a mathematical point of contact. In the most general case this might mean a retreat all the way back to set theory, but in the present example it would mean a causal progression in a timelike direction. We can then fantasize using this mathematical connection to somehow tunnel across the ultimately intimidating space of all mathematical constructions—perhaps through a dragon hole, named after a creature with the same magical power and current degree of reality as the level 4 multiverse.

If we back away from the statement at the top of this subsection to a *limited* set of alternative mathematical possibilities, we are contemplating a 4- universe which requires less stretching of credibility. We only have to accept that our single spacetime manifold is not alone in the entire expanse of all that exists.

**12.5. The progressive enlargement of our worldview**

Here we have argued that the level 1- and 3 multiverses have become a proper part of science, because they are implied by observation, experiment, and logic. For example, cosmology in the level 1- description has now become a thoroughly convincing and quantitative science.

The same is true of quantum theory in the level 3 description, as demonstrated by increasingly precise quantitative tests and increasingly sophisticated demonstrations of entanglement at a distance ([187] and references therein). At this point, in fact, a resistance to the natural Everett framework in thinking about quantum mechanics might be a mild impediment in developing quantum technologies for communication, computing, etc., in the same way that resistance to the theory of evolution can be an impediment to developing biological and medical technologies. In both cases a clean way of thinking can be more effective than one impeded by philosophical reservations. One expects that a deeper theory underlying current quantum physics will eventually be discovered, but the successful basic predictions of quantum physics must still hold up, since they have been so well tested—in the same way that the description of planetary motion by Newtonian dynamics survived the deeper theory of Einstein.

The level 1 and 2 multiverses have some plausibility for the reasons given above and in much more extensive treatments, including those cited here.

The level 1+ multiverse is worthy of consideration because we do not know if our full spacetime manifold has positive, zero, or negative curvature, and in the last two cases it has infinite extent.

In entering the level 4−multiverse, we finally leave our own (D-dimensional) manifold and envision that there is more to the entire extent of Nature. And if we are still bolder, we can entertain the thought of Tegmark's far-reaching Mathematical Universe Hypothesis and the resulting level 4 multiverse.

Over the years there have been fears of invasion from another planet, as in Orson Welles' 'War of the Worlds' radio broadcast in 1938 which frightened hundreds of thousands of people. What is the possibility of an invasion from another *universe*? The discussions above imply that there may be a 'universe protection principle', resulting from the fact that the physics required is far beyond anything we can currently imagine. For example, an attempt to traverse even the type 1- multiverse appears to require the implantation of both ends of the required wormhole in the required locations—one near us and the other near the distant aliens—before they are separated by cosmic expansion. Similarly, type 2 universes have been separated since the Big Bang—or have always been separated—and type 3 universes have been separated since the moment of decoherence. Type 4 invaders would find our world quite inhospitable (even more so than those from a different type 2 universe), and they would find the journey even more difficult. It is probably also safe to assume that beings with unimaginably advanced technologies will be above trivial territorial ambitions.

As science enlarges our view of Nature, there is often an emotional back reaction, as in Walt Whitman's 'When I Heard the Learn'd Astronomer':

*When I heard the learn*'d astronomer,
*When the proofs, the figures, were ranged in columns before me*
*When I was shown the charts and diagrams, to add, divide, and measure them,*
*When I sitting heard the astronomer where he lectured with much applause in the lecture-room,*

> *How soon unaccountable I became tired and sick,*
> *Till rising and gliding out I wander'd off by myself,*
> *In the mystical moist night-air, and from time to time,*
> *Look'd up in perfect silence at the stars.*

The appropriate response is from Feynman:

> Poets say science takes away from the beauty of the stars—mere globs of gas atoms. I too can see the stars on a desert night, and feel them. But do I see less or more? The vastness of the heavens stretches my imagination—stuck on this carousel my little eye can catch one-million-year-old light. A vast pattern—of which I am a part... What is the pattern, or the meaning, or the why? It does not do harm to the mystery to know a little about it. For far more marvelous is the truth than any artists of the past imagined it.

Whitman was a better master of language, but Feynman had the wiser perspective. For those with emotional reactions against the modern scientific worldview, it should be emphasized that we are enhanced rather than diminished. All the past and future revelations about the full scale of Nature, and our own place in Nature, should lift our spirits and enrich our lives. And we should not be any more uncomfortable with quantum reality than we are with the fact that we are rapidly moving through space, by more than 200 kilometers every second, or the fact that our remote ancestors were one-celled creatures—facts which many first rejected as absurd.

Accepting quantum mechanics *per se* without philosophical boilerplate or angst (i.e. the Everett interpretation), or, when required, accepting other aspects of a multiverse, has no implications for daily human behavior, in the same way that acceptance of biological evolution, the implications of neuroscience for human consciousness and free will, etc will not have direct impact on how we live. But, in the long run, we will benefit from a worldview that is logically and scientifically consistent, free of fuzzy thinking and intellectual dishonesty.

Please note that in figure 6 Bohr and Everett are smiling at one another. Let us continue this tradition with tolerance for those who, at the moment, have differing points of view. According to the Everett interpretation, the answer to the question of Ed Fry earlier in this paper is that the photon approaching a filter merely heeds the injunction of Yogi Berra, 'When you come to a fork in the road, take it'. But we should also remember that Ed has followed in the tradition of a long line of distinguished physicists like Richard Feynman, who once said 'Nobody understands quantum mechanics'.

## 13. Towards a machine that works like the brain: the neuromorphic computer by Ivan K Schuller, Sharon Franks, Oleg Shpyrko and Alex Frano

'Moore's law', the doubling of computational power every year and a half, has fueled the large explosion in the use and manipulation of data in our everyday lives. However, it is widely agreed that in the next two decades a 'Moore's crisis' will develop, in which the continuous improvement in computational power and the exponential decrease in cost will slow down dramatically. For this reason, a worldwide quest to find new computational paradigms is underway. 'Neuromorphic computing' refers to a scientific aspiration to develop a computer that works like the human brain (see figure 10). Why this aspiration? What makes it so challenging? What is the role of physics in this ambitious undertaking?

We humans are generating and using data at ever-increasing rates. Yet current technologies can no longer keep pace with society's ever-growing computational needs. We will need to develop entirely new types of computers that work differently—and far more efficiently [188]—than those we use today. This is where we can look to the human brain for inspiration.

Our brains are not only capable of rapidly deriving meaning from complex inputs, they do so using remarkably little power. The brain is extraordinarily energy efficient. Any viable solution to meeting projected demands for computational power will need to be energy efficient. Why the need for greater energy efficiency? The principal reason energy efficiency is vital is because conventional, energy-hungry computers generate vast amounts of heat [189]. If not dissipated, the heat interferes with the functioning of the computer. Using current technology, to build a reasonably-sized computer—for example, one that is not the size of an aircraft carrier—with capabilities of the human brain, would require packing the hardware so closely that the heat-dissipation challenge would be insurmountable. An additional obstacle is that producing, manipulating, and storing large amounts of data consumes vast amounts of energy. It is estimated that approximately 10% of the world's energy consumption is now used for data manipulation. Energy use for data manipulation continues to increase with time, as society's demand for computational power seems to be insatiable.

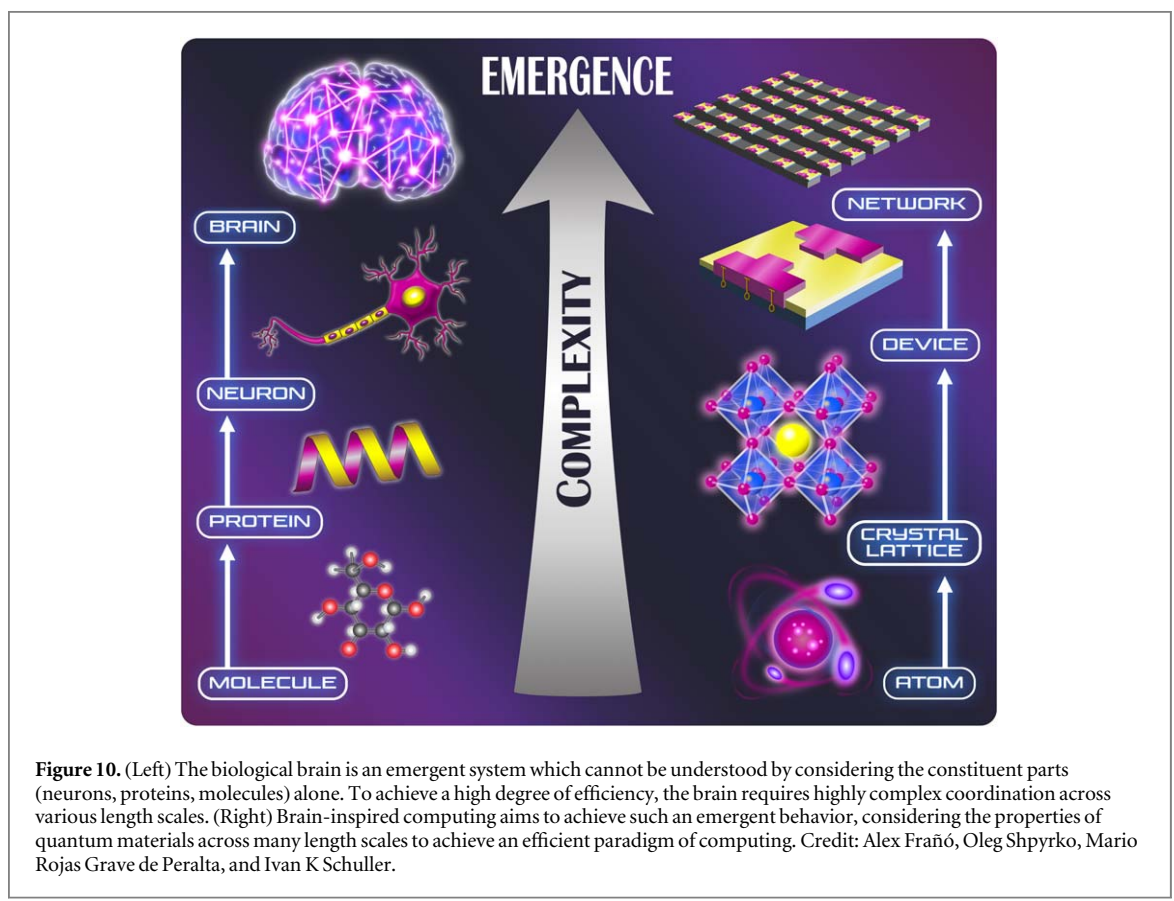

**Figure 10.** (Left) The biological brain is an emergent system which cannot be understood by considering the constituent parts (neurons, proteins, molecules) alone. To achieve a high degree of efficiency, the brain requires highly complex coordination across various length scales. (Right) Brain-inspired computing aims to achieve such an emergent behavior, considering the properties of quantum materials across many length scales to achieve an efficient paradigm of computing. Credit: Alex Frañó, Oleg Shpyrko, Mario Rojas Grave de Peralta, and Ivan K Schuller.

In order to build a neuromorphic computer, we need devices that function like the brain's synapses, neurons, axons, and dendrites. In fact, such artificial brain components already exist using conventional hardware [190, 191]. The challenge of making these hardware constructs as energy efficient as the brain still looms.

The human brain requires only about 20 watts and contains approximately $10^{11}$ neurons and $10^{14}$ synapses, so it requires a piddling 0.1 picowatts ($10^{-13}$ W) per synapse. To achieve comparable energy efficiency in a computer, devices based on entirely new 'quantum materials' are showing promise.[24] These new materials enable behaviors and functionalities—non-linear, tunable processes that we need to understand, and eventually control and exploit. Such control is particularly important at nanoscale dimensions, where non-linear behavior can induce high thermal gradients that push materials very far from equilibrium. Understanding and controlling the behavior and thermodynamics of nanoscale materials and devices far from equilibrium is where physicists are poised to make paradigm-shifting contributions. Ultimately, the realization of neuromorphic computing as a disruptive technology will require intensive, sustained collaboration among physicists, materials scientists, device designers, systems engineers, computer experts, mathematicians, biologists and neuroscientists. Inspiration from biological systems, combined with scientific innovation, and fueled by the engagement of creative minds may one day fulfill the dream of developing a machine that works like the human brain [192–198].

## 14. What can we say about the 'Value of information' in Biophysics? by Lázaro A M Castanedo, Peyman Fahimi and Chérif F Matta

Herein follows a flavour of a few seemingly fertile ideas from a voluminous literature of potential import in the development of biophysics. We demonstrate how some aspects of a theory developed by engineers to address problems in communication engineering are transferable to the realm of biology. Might *the specific problems of biology* return the favour one day, suggesting an extension of the classical theory of communication.

---

[24] For further information on a project dedicated to the development of Quantum Materials for Energy Efficient Neuromorphic Computing see http://qmeenc.ucsd.edu.

### 14.1. Early hints for a central role of 'Information' in biology

Modern biology and biochemistry textbooks abound with phrases like *genetic code*, *genetic message*, *genetic information*, *replication*, *transcription,* and *translation* reflecting biology's celebrated central dogma [199], that genetic information is passed unidirectionally from DNA to RNA to protein. A mutation is a *change in the genetic information* or an error in the *copying* of this information, either spontaneously or as a result of interaction with radiation, mutagens, or viruses. These information-theoretic sounding phrases can be traced-back to Erwin Schrödinger's influential monograph *What is Life?* [200] in a section titled *The Hereditary Code-Script (Chromosomes)*.

In 1928, Frederick Griffith discovered that dead *pneumococci* carry a substance he termed a *transforming principle* that is able to transmit *heritable* virulence in non-virulent strains of the live bacteria. Ironically, Schrödinger's book appeared in the same year (1944) as the definitive paper by Oswald Avery, Colin MacLeod, and Maclyn McCarty establishing DNA as the *transforming principle* and, hence, that DNA is the physical carrier of the genes [201]. Remarkably, however, the book predates by almost a decade the first reports of the discovery of the double helical structure of the DNA polymer by James Watson and Francis Crick [202], and simultaneously—by Rosalind Franklin, Raymond Gosling [203] (and Maurice Wilkins) –the discovery that suggested an actual implementation of a code-like mode of operation for DNA [204–206]. Just a year later, the direct correspondence between the DNA language and its protein translation was proposed by the Russian physicist George Gamow [207] (although the details of how this is achieved are now known to be different that Gamow's lock-and-key proposition).

Schrödinger concludes this section by describing chromosomes with the words: '[*t*]*hey are law-code and executive power or, to use another simile, they are architect*'s *plan and builder's craft in one*' [200]. The brilliant experiments of Leonard Adleman in the 1990's showed how DNA can be programmed into actual software to solve the traveling salesman problem numerically in the test-tube [208].

Today, following a terminology that appears to have been coined by Michael Polanyi, the distinguished physical chemist and philosopher, DNA is often referred to as the *blueprint* of life [209]. But a blueprint is essentially condensed *information* with the *potential* to give rise to a physical object if executed. It need not even be complete since the code's implementation interacts with the environment in producing the resulting individual, as captured by the popular phrase '*Nature and nurture*'.

### 14.2. The quantity of information stored in nucleic acids and proteins: syntax

Coincidentally, the end of the 1940s also saw the birth of Claude Shannon's (classical) 'Information Theory' [210, 211], a theory originally conceived in an engineering context to optimize the transmission of information through electrical wires. It did not take long for scientists to realize the relevance of this nascent theory to the realm of biology [212–218].

The intellectual atmosphere that catalyzed this appropriation was, perhaps, epitomized by the position of Michael Polanyi, who has argued very strongly against a strong reductionist approach to biology. Polanyi was simply not convinced of the possibility, even in principle, of reducing biology to chemistry and then to physics (classical electromagnetic theory and quantum mechanics), where each lower level represents a 'more fundamental' underlying level of description [209, 219]. For Polanyi, a living system is analogous to a 'machine' in many respects, i.e. to a 'mechanism' that operates in full compliance with the laws of physics and chemistry but within 'boundary conditions' that are in themselves *not reducible* to such laws (despite not violating them) [219].

After arguing that a watch, for example, is more than just the atoms that compose it since its design as a functioning time measuring device is *not* a consequence of the laws of physics, Polanyi transits to biology by the following revealing statement: [219]

> *Now, from machines let us pass on to books and other means of communication. Nothing is said about the content of a book by its physical-chemical topography. All objects conveying information are irreducible to the terms of physics and chemistry.*

Clearly a theory of biology should somehow incorporate information theory since important aspects of its essence are boundary conditions that cannot be reduced to the laws of physics and chemistry. Biopolymers such as DNA, RNA, or proteins are a case in point. The sequence of the monomers composing those polymers is 'dictated' over millions of years of evolution by unknown environmental factors and is now a 'given', intrinsic to the individual from the start of its existence. The elevated temperatures at which biological systems operate will quickly destroy any quantum coherence of entangled quantum states [220] leaving classical information theory [210] as the appropriate framework within which to study biological information. Before proceeding further, a clarification is needed. While chemical composition is irrelevant for the intended operation of a watch, in the case of the DNA, chemical structure is indispensable for its function (otherwise, for instance, how could DNA be a good substrate for the DNA polymerase or transcriptase?). The watch-DNA analogy is only meant to underscore that the *information* carried by the genetic material is independent of the underlying substrate.

Influenced by Polanyi's philosophy, Lila Gatlin wrote her classic monograph '*Information Theory and the Living System*' [213]. The physical transmission of information from a source (e.g. DNA) to a recipient (e.g. the ribosome and eventually a protein, via mRNA) is accompanied by 'noise' which may result in loss or destruction of some information, that is, an increase in the entropy of the message. A machine such as a living cell can minimize noise by ensuring that the message to be transmitted has excess information, with effective repetition providing redundancy. Chargaff's rules [221], predating the discovery of the double helix, stipulate that the composition of DNA must have equimolar amounts of the complementary bases, so [A]=[T], and [G]=[C], where A is adenine, T is thymine, G is guanine and, C is cytosine, and where the square brackets denote molar concentrations of a given base. (DNA would come to explain this through the Watson-Crick hydrogen-bonding complementarity rules, whereby A must bind to T and G to C.) However, there are no rules regulating the proportions of the AT pair with the CG pair. Thus, in real DNA, the composition is such that the concentration of AT and GC are generally different ([AT] $\neq$ [GC]), and the proportion ([AT]/[GC]) characterizes the specific organism.

For a language consisting of $N$ symbols, Shannon's average information content per symbol in the message is given by the well-known relation: [210]

$$H_1 = -K \sum_{i=1}^{N} p_i \log_2 p_i, \qquad (2)$$

where if $K = 1$ and is dimensionless, $H_1$ is in bits (the unit adopted here), and the subscript '1' denotes that this is the average information per symbol. In equation (2), if $K = k_B \ln 2$ (where $k_B$ is Boltzman's constant) then $H_1$ is in units of entropy—which actually connects physical entropy and information. It is worth noting that the common dimensionless unit of information, the bit, introduced above is short for 'binary digit'. If the logarithm to base 10 is the one inserted in equation (2), the unit is termed the 'hartley or Hart' in honor of Ralph Hartley, or the 'dit' meaning decimal digit. Finally, if the natural logarithm ($\ln \equiv \log_e$) is used, the unit is the 'nat', i.e. the natural unit of information.

To maximize $H_1$ one has to equalize the probabilities of all the symbols $\{p_i\}$. The greater the departure from equiprobability, the smaller the information content of the message (to the extreme case where one symbol has a probability of 1 and all the rest zero probability, and no information is conveyed at all). In the English language, for example, the 26 letters appear with different frequencies, with 'e' being the most common (probability ~ 11%) and 'q' the least probable (0.2%) where normalized frequencies are considered probabilities. English, therefore, has a lower information content per symbol than an 'ideal' language would have with all the letters being equiprobable ($p=1/26$). For an organism with the unlikely equiprobability of the four nucleobases, i.e. with [A]=[T]=[G]=[C]=25%, as in *E. coli*, $H_1 = -\log_2 0.25 = 2$ bits per symbol. This is the maximal information carrying capacity of a nucleic acid base.

The departure from equiprobability of base pairs means that the probabilities of each of the four individual bases in the genome differ from the ideal value of 1/4. Gatlin defined the redundancy in the genetic message due to a *departure from equiprobability* as: [213, 214].

$$D_1 \equiv H_1^{\text{Max}} - H_1^{\text{Actual}}, \qquad (3)$$

with the subscript '1' indicating this first 'type' of redundancy.

Gatlin then defined a second type of redundancy, exhibited in the genome, the *departure from independence*. To illustrate what this means, let's return again to the structure of the English language where, for example, the letter 'q' is followed by 'u' with high probability (e.g. e*qu*al, *qu*ality, or e*qu*iprobable)—thus the appearance of a given letter depends on the previous one. This is termed a first order Markov process (although higher orders of Markov processes exist, we limit ourselves to the first order for simplicity). Such a Markov process constitutes redundancy since it decreases the freedom of choice of symbols.

In the absence of this second type of redundancy, the information content of an n-tuple (a block of $n$ symbols) is expressed by [213, 214]:

$$H_n^{\text{ind}} = -\sum_{i=1}^{N}\sum_{j=1}^{N}\cdots\sum_{n=1}^{N} p_i p_j \ldots p_n \log_2(p_i p_j \ldots p_n) = nH_1, \qquad (4)$$

implying that the total information content of the $n$-tuple is nothing but $n$ times the average information content per letter or symbol.

Generally, however, there will be departures from this independence. Limiting the discussion to a first order Markov source (with a memory $m=1$), where the probability of a given letter in the message depends only on the letter immediately preceding it in the sequence, with a departure from independence the information content of the n-tuple is now [213, 214]:

$$H_n^{\text{dep}} = -\sum_{i=1}^{N}\sum_{j=1}^{N}\cdots\sum_{n=1}^{N} p_i p_{ij} \cdots p_{(n-1)n} \log_2(p_i p_{ij} \cdots p_{(n-1)n}), \quad (5)$$

where $p_{ij}$ is the probability of appearance of the *j*th letter given that the previous letter in the message is *i*. With some manipulations, equations (3)–(5) lead to [213, 214]:

$$D_2 \equiv H_2^{\text{indep}} - H_2^{\text{dep}} = H_1 - H_{\text{Markov}}, \quad (6)$$

where:

$$H_{\text{Markov}} = -\sum_{i=1}^{N}\sum_{j=1}^{N} p_i p_{ij} \log_2(p_{ij}). \quad (7)$$

The total redundancy in a DNA sequence is then:

$$R \equiv \frac{D_1 + D_2}{\log_2 4} = 1 - \frac{H_1^{\text{actual}}}{H_1^{\text{ideal}}}, \quad (8)$$

where 'actual' refers to the characteristic redundancy of the chosen language and ideal refers to a language using the same letters but with all letters equiprobable and independent.

Redundancy measures the constraints imposed by the structure of the language that are designed to reduce transmission errors in a message expressed in that language. It is conceivable that one of the measures of evolutionary 'fitness' is how well an organism has maximized *R* while keeping the genetic language sufficiently flexible to code for its enormously complex structure. Since there is an inverse correlation between the redundancy and the number of potential messages expressible in a given number of symbols, a compromise must be struck.

Gatlin noted that at different steps in the evolutionary ladder organisms achieve this (constrained) maximization of redundancy by different means. The higher the organism in the evolutionary tree, the more it achieves a higher *R* by keeping $D_1$ relatively constant while maximizing $D_2$. The converse is true for lower organisms which maximize their redundancy mainly by maximizing $D_1$. An enormous body of literature took these ideas as a point of departure in the final decades of last century to classify organisms, quantify differences between sequences, compare coding and non-coding regions of DNA, and compare homologous sequences from different organisms [215, 222]. All these ideas that apply for nucleic acid also apply for proteins, but with an alphabet comprised of 20 amino acids, which if they were equiprobable and independent would transmit a maximum of $\log_2 20 = 4.322$ bits per amino acid.

Exciting as it may be, the application of Shannon ideas to nucleic acids and proteins is limited in a significant and fundamental way—information content is a measure of entropy, no more.

### 14.3. The value of information stored in nucleic acids and proteins: semantics

Mikhail Volkenstein stressed the limitations of information content/entropy, emphasizing instead how one must consider the *value* of information in biology, instead of taking only the *quantity* of information into account) [223–226]. Shannon's theory quantifies the amount of information (number of bits) in a message, but says nothing about the importance of this information. Volkenstein describes [the eminent Soviet evolutionary biologist] Ivan Schmalhausen's pertinent remark that

> the current information theory has no techniques available to it for evaluating the quality of information, although this factor is often of decisive importance in biology. When an organism receives information from the environment, first of all it evaluates this information from the standpoint of its quality… as 'irrefutable' [227]. This statement remains essentially true today, and it is a task for the future to construct a theory of the value of biological information starting, perhaps, from where Volkenstein left off (*vide infra*).

Volkenstein realized that the effect on a recipient receiving information is a measure of the value of the information. He exemplified this with a 'fair traffic light', meaning one that is red and green for equal amounts of time. The emission of one bit of colour information would cause considerably greater traffic to flow on a large avenue than on a small side street. Thus identical information in the Shannon sense can have dramatically different consequences depending on the receiving system [227].

Volkenstein relates the value of information to its irreplaceability, that is, non-redundancy. He argues further that the value of the information increases gradually in the course of evolution. He gives the following intriguing definition of the (dimensionless) value of information as [223, 225]:

$$V = \log_2\left(\frac{P_\text{final}}{P_\text{initial}}\right), \tag{9}$$

where $P_{initial}$ and $P_{final}$ are the probabilities of producing a given effect or outcome before and after the receipt of information by the receiving system. (See [225] and references therein for the justification of choosing this definition.) A reasonable 'target' for an organism is to live as long as possible, while the 'goal' of DNA is eventual protein synthesis.

New information is generated every time an individual of any given species is conceived through sexual reproduction by receiving half of its genetic material from its mother and half from its father. The act of sexual reproduction includes a series of random events that are not easily traceable to the laws of physics and chemistry, e.g. the decision of a particular male and female to mate. The selection of a mate can be regarded as Polanyi's 'boundary condition', [209, 219] untraceable to (but of course not violating) the laws of physics and chemistry (*vide supra*).

The form of equation (9) allows for positive or negative values of information. Imagine, for instance, that a professor, after spending an hour in class deriving an equation, discovers a mistake at the very beginning of the derivation and closes the lecture by informing the students that the entire derivation was wrong.[25] This last piece of information invalidates all information passed on during the class, and hence, has a negative value. Value can also be a function of time. Information about an impending attack by the enemy's army is actionable intelligence (of high value) *before* the attack but worthless once it has happened. Further, repetition of the message before the attack has no value—it is totally redundant.

Let us examine how redundancy plays out in the eventual translation of a DNA message in a protein coding gene into the corresponding protein, assuming equiprobability of symbols for simplicity. First, in passing, we recast the triplet nature of genetic codons (the three nucleic acid base letters per amino acid) in terms of information theory. This minimal number of nucleotides per amino acid emerges from the ratio of the maximum information per letter of protein divided by the minimum information per letter of DNA, i.e. 4.322/2.000 = 2.161 which, as there are no fractional nucleotides, necessitates three nucleotides per amino acid.

Now, for a protein-coding gene containing $n$ equiprobable nucleotides, $H_1^{DNA} = n \log_2 4 = 2n$ bits. When translated to a protein, this will correspond to $H_1^{protein} = \frac{n}{3} \log_2 20 = 1.44n$ bits, i.e. there is a compression of the information on passing from DNA → protein at even at the most basic level where all bases and amino acids are equiprobable. In other words, a redundancy of 1 −1.44/2.00 = 0.28 exists in the primary sequence of DNA gauged with respect to its protein translation, owing to the degeneracy of the genetic code. Hence there is an increase in the value of information at the protein level –under these idealized conditions –compared to the value in the DNA sequence.

On the other hand, non-redundant information is irreplaceable. Here is where the definition in equation (9) comes into play. Take for example a point mutation (i.e. a mutation that changes the nature of only one of the three symbols (*x, y, z*) in a codon). If this mutation results in a significant change in the hydrophobicity/hydrophilicity of the coded amino acid (measured by free energy of transfer from a polar to a non-polar medium or to the gas-phase) [228–230] then this mutation is poised to have drastic effects on the protein's overall three-dimensional structure. The value of the information replaced by this mutation is, consequently, high.

The degeneracy of the genetic code is primarily in position *z*, in other words, synonymous codons (codons coding for the same amino acid) usually differ in the third position, and hence the *z*-position is the least important (least valuable) position of a given codon. Meanwhile, the middle letter, *y*, determines whether the coded amino acid is hydrophobic or hydrophylic [228]: It is hydrophobic if this letter is of the pyrimidine family (C or U) in the mRNA codon and hydrophilic if it is of the purine family (G or A). Furthermore, the middle letter is *unique* for a given amino acid (except for serine in which it could be either G or C), hence a mutation in the *y*-position almost always changes the amino acid. Thus, this letter is the most valuable since it is likely to have the most drastic consequence on the ensuing protein structure. Nature has fine-tuned the code in such a manner that the probability of replacing a residue by one with different hydrophobicity is minimized [224]. Degeneracy plays a much wider role in biology as argued forcefully in an important review by Edelman and Gally [231]. In this review, the authors provide a tabulation of the degeneracy at 22 different levels of hierarchical organization in biological systems e.g. molecular (as the degeneracy of the genetic code), macromolecular (proteins with very different primary structure that can assume similar overall morphology and function), up to the macroscopic level of human and animal communication (see table 1 of [231]). In fact it is the very presence of degeneracy that provides the 'raw material' for natural selection and evolution [231].

Alternatively, one can define the value of amino acids as measured by their degree of conservability (irreplaceability) in homologous protein from different species since conserved residues are presumed of higher value. Originally, Volkenstein relied on Dayhoff's matrices of amino acid replaceability in defining the value of a given amino acid, following Bachinsky, where the 'functional similarity of amino acid residues (FSA)' is defined as [224]:

$$\text{FSA} = \left(\frac{2N_{ij}}{N_i + N_j}\right), \quad (10)$$

where $N_{ij}$ is the number of times amino acid *i* is replaced by amino acid *j* within a set of homologous proteins, and where $N_{i,j}$ are the abundance of the *i*th or *j*th amino acid in the given set, respectively. The resulting (non-symmetric) matrices are 21 × 21 in size (20 amino acids + a termination code). They are non-symmetric because the propensity to replace (mutate) amino acid *i* by *j* is not generally equal to the probability of replacing *j* by *i* in the course of evolution.

---

[25] This example is not original, it was read or heard by one of the authors (C.F.M.) who regrets that he is unable to recall the source to cite it.

Using these matrices and definition (10), Volkenstein then estimates the FSA for every possible single-point mutation of every codon of the 64 codons of the genetic code. A code *x,y,z* can have 9 single point mutants (since we have 4 bases, one of which is already used, so the possible mutants are 3 per position×3 positions). If a single point mutation of a codon coincides with the same amino acid, a silent mutation, it is arbitrarily given an FSA=100. The nine FSAs for every codon are then averaged (and divided by a numerical constant to retain a convenient magnitude), yielding *q*, defined as a measure of the codon irreplaceability. The value *v* of a residue is greater for smaller *q*. As an example, say the codon AAA (for lysine), yields $q = 0.74$. The value of this codon is then $v = (q + \frac{1}{2})^{-1} = 0.81$. Proceeding in this manner for all 61 unique *x,y,z* sense codons, the result is a genetic code table with a numerical value assigned for every coding codon [224].

If we now average the values of the $(x_i, y_i, z_i)$ degenerate codons (different codons coding for the same amino acid), we get the value of the coded amino acid in a protein. (See table 9.3, p. 264, of Volume II of [224] ). Accordingly, the most valuable (the most irreplaceable) amino acid is tryptophan ($v_{\text{Trp}} = 1.82$) and the least valuable is alanine ($v_{\text{Ala}} = 0.52$) [224, 232]. Curiously, we note here in passing, that the partial molar volume as well as the quantum mechanically calculated molecular volume of the hydrogen-capped Trp side-chain happen to be the largest among all 20 amino acids, while that/those of Ala are the smallest, [229, 230] a coincidence perhaps, but possibly worth exploring.

The average *changes* in the hydrophobicities of amino acids resulting from replacements of the type $x \rightarrow x'$ and $y \rightarrow y'$ indicate that the 'least dangerous' mutation is of the type A ↔ G [232]. While there is a wealth of fascinating findings that we skip in this brief essay, one that stands out is that evolutionarily older proteins such as cytochrome c, unlike much more recent ones such as hemoglobin, tend to have a higher value in species that are higher in the taxonomical tree, with humans at the very top [224].

### 14.4. Closing remarks

Cannarozzi *et al* [233] re-evaluated some of the measures of irreplaceability described above using the much larger and more recent database of Jiménez-Montaño and He [234]. In doing so, Cannarozzi *et al* [233] obtain an agreement of ∼87% in the calculated values proposed by Volkenstein who used a smaller and older database [224]. Thus it appears that Volkenstein's core ideas are essentially correct even on quantitative grounds. But the field would benefit from a revisit using the most up-to-date and extensive data and from the formulation of a full and consistent *theory of the value of biological information*, a theory that can serve both biophysics and communication engineering.

Today, in 2023, our knowledge has soared to unprecedented heights. That the entire human genome has been sequenced [235] is already considered history, not to mention the sequencing of the full genomes of dozens of other species. Bioinformatics is a mature field [236, 237]. UniProt [238, 239] annotates more than 20,000 proteins and their properties and locations of their coding genes. It is well established that only 2% of the genome consists of protein coding sequences while the rest of the genome does not code for any protein (non-coding DNA, or ncDNA). Non-coding DNA represents the bulk of nuclear DNA (98%), and its functions in living cells—if any—remain essentially an open problem. What would be the effect of mutation on these ncDNA sequences and what is their role in the first place? Are there information theoretic differences between coding and non-coding DNA? Can information theory shed light on the function of repetitive DNA segments (half of the human genome) such as tandem repeats of trinucleotides and their roles in genetic diseases such as Huntington's disease [240]? Are there information theoretic differences between nuclear and mitochondrial DNA? What is the effect of ncRNA on the rate of the translation step (its kinetics) which can affect protein folding through translational pausing?

Irreplaceable (high value) amino acids must be crucial for the function of the protein and, hence, obvious targets for drug design and for manipulations by site-directed mutagenesis and/or *in vitro* directed evolution and for understanding genetic disorders and viral and bacterial development of resistance (see [234] and references therein). It is entirely possible that equation (9) is an over-simplification, which invites further investigation into the meaning of the value of information. Might this ultimately lead to new physical theory, or perhaps even a sub-branch of the mathematics of communication?

But the role of information theory in biology does not stop at analyzing sequences. Information itself is physical, as Landauer taught us long ago [241], and to erase it energy must be expended. The energy to erase one bit is small ($k_B T \log 2$), but if this erasure is repeated by a molecular machine at a high turnover rate, the informational cost starts to be consequential. The old paradox of the extreme inefficiency of the kidney compared to any other bodily organ can only be resolved by accounting for the information theoretic cost of recognizing ions e.g. $Na^+$ to be selected and sorted for excretion by the kidney [216–218]. These ideas also place a limit on the thermodynamic efficiency of a molecular machine like ATP synthase/ATPase which acts as a sorting machine, picking protons for transport *parallel* or *antiparallel* to a pH gradient, respectively, across mitochondrial inner membranes or bacterial membranes [242–246].

Interesting problems that do not appear to have been explored (at least extensively) in the literature include the reformulation of the following type of engineering problems into a biological context: *Packet loss* (i.e. the failure of a message to reach its intended destination); *bit rate* (the *rate* of information transmission); *transmission delays* (the time needed for a signal to flow in its entirety through a communication channel).

Translational pausing during translation regulates the rate of information flow through the mRNA-ribosome informational system apparently to allow the nascent protein sufficient time to fold properly. How is the pausing coded in the mRNA message? It is tempting to think of the information coded in the mRNA as having a dimension greater than one, where the extra dimension regulates the rate of translation.

Another issue concerns the exploration of other definitions of classical information such as the Fisher information [247], originally proposed in 1922 (before Shannon's definition). Shannon's information is a 'global' measure since it involves a summation (and, in the limit, an integration) over the entire message. In contrast, Fisher information involves an integration over the *gradient* of the probability distribution function, and hence is sensitive to and magnifies local variations in the probability distribution function [247]. Can Fisher information play a role in pinpointing hot-spots in biological messages?

In closing, we draw the attention of the reader to a 1991 commentary by John Maddox '*Is Darwinism a thermodynamic necessity?*' [248] on the then recent paper by J-L Torres in which the latter proposes a thermodynamic formulation of the ill-defined concept of Darwinian 'fitness' [249]. The purpose of the highlighted paper is to translate Darwinian's 'fitness' into quantitative deviations from a set of ideal thermodynamics parameters characterizing a living system. Torres has succeeded, at least in principle, in lifting the circularity of the 'survival of the survivors (fittest)' [249]. Could the '*value*' of a nucleic acid or a protein be an alternative, or perhaps an additional or complementary, measure of the fitness of a species from an evolutionary standpoint?

## 15. What breathes the fire of consciousness into our brains? by Suzy Lidström and Solange Cantanhede

It is remarkable that human consciousness, long regarded as an immaterial or even spiritual phenomenon, is increasingly revealed to be associated with well-defined physical processes in the brain (see [250] and also, for example, the more recent [251–253] and references therein). There are three timescales associated with consciousness: the first is the moment-to-moment experience of conscious awareness. The second is the growth of consciousness from a single cell into an organism with trillions of cells. The third is the evolution of consciousness in the biosphere over millions of years. Each has an analogue in physics. We have interpreted conscious processes on the timescale of seconds as the coherent excitation of quantum fields, analogues to collective modes in condensed matter physics [254]—for example the hybrid modes of the electromagnetic field and electrons in an ionic crystal. The growth of a conscious brain is a vastly more sophisticated analogue of the growth of crystals or other ordered phases, and the evolution of consciousness is crudely analogous to the evolution of contemporary quantum fields from other more primitive quantum fields of the early universe.

### 15.1. Two perspectives on the brain—a biased history

The experimental work of Nobel Laureate Santiago Ramon y Cajal [255], including the complex (and beautiful) drawings of neurons [256], spearheaded close to one century of research dedicated to unravelling the inner workings of the human brain. Research addressed different scales, from the fine detail of the operation of individual neurons, to consideration of the billions of neurons concentrated in the outer few millimeters of the cerebral cortex, and, taking a still broader brush, investigations of the electrical signals in the brain [257, 258]. The resultant understanding of the grey matter [GM] is such that, for example, the transmission of a single spike can now be described in detail as it journeys through the brain [259], and the activity in specific neurons of the brain can be associated with particular thought processes or actions (such as the fusiform face area, a key breakthrough towards the end of the last century [260]). In a clinical setting, routine treatments exist. One such is deep brain stimulation, which is applied to suppress the tremors associated with Parkinson's disease and, with the assistance of MRI images and connectomics targeting, to provide relief for patients with depression; it is also being assessed for numerous other applications (e.g. [261]).

As the understanding of the GM grew, the tremendous import of the white matter [WM] became apparent [262]: the conductive properties of neurons are enhanced and modified by myelin, a fatty insulator which encircles the axons in sheaths that are broken by gaps along the length of the axon, enabling the glial cells and oligodendrocytes (that also comprise the white matter) to perform tasks such as alimentation, repair, and alteration [263].

As almost half of the adult human brain is comprised of white matter, giving us 20% more white matter than chimpanzees and a massive 500% more than mice [264], the fact that it conveys an evolutionary advantage should come as no surprise (see e.g. [265] and [266] on the evolution of the human brain). Indeed, although sustained efforts have been made to teach primates to communicate in diverse ways, the limitations of these studies are perhaps more telling than the successes: After years of intensive one-to-one tuition, primate brains have a measurably thicker cortex than that of members of their species not subjected to an intense learning regimen, yet their achievements pale into insignificance when compared with human learning over the same period of time. We can only acknowledge that the human brain has an astonishing ability to learn. This ability escalates when motivation is high, a truth captured by William Butler Yeats when he said: 'Education is not the filling of a pail, but the lighting of a fire.'

Once perceived as little more than biological scaffolding and electrical insulation, the sheaths of myelin encasing neurons, the astrocytes and the glial cells have been recognised to be vital:

- for cognition, behaviour, development, and learning (see [267] and references therein), including the attainment of expertise [268, 269];

- to achieve fully fledged brain function including the optimal development of executive functions (see, e.g. [262, 264, 267, 269] and references therein);

- for the brain's plasticity [270]; and

- in the central nervous system (see, e.g. [271]).

In addition to increasing the velocity of action potentials, as action potentials themselves affect local protein synthesis and myelination, reciprocal fine-tuning of spike transmission and enhanced synchronisation result [258, 262, 272–274] and [275]. White matter facilitates connectivity through axons of various kinds, enabling clusters of neurons in different, and sometimes widely separated, regions of the brain to act in synchrony. The frontal lobes, which have an 'abundance' of white matter: '… have the highest degree of connectivity of any brain lobe.' [263]

When we refer to the *grey* matter, though, we need to realise that 'there is no GM in adult humans without substantial amounts of myelin in it' [262]. Pease-Raissi and Chan [275] refer to the '(w)rapport between neurons and oligodendroglia', clarified as the 'reciprocal relationship in which neurons alter oligodendroglial form and oligodendrocytes conversely modulate neuronal function.' Their review summarises the advances in our understanding of the role myelin plays, and outlines important ongoing research areas:

> Myelin, multilayered lipid-rich membrane extensions formed by oligodendrocytes around neuronal axons, is essential for fast and efficient action potential propagation in the central nervous system. Initially thought to be a static and immutable process, myelination is now appreciated to be a dynamic process capable of responding to and modulating neuronal function throughout life. While the importance of this type of plasticity, called adaptive myelination, is now well accepted, we are only beginning to understand the underlying cellular and molecular mechanisms by which neurons communicate experience-driven circuit activation to oligodendroglia and precisely how changes in oligodendrocytes and their myelin refine neuronal function.
>
> *Pease-Raissi and Chan, 2021*

The combination of electrical and chemical processes involved in signal transmission has transformed our understanding of the neuron from that of a conventional passive conductor—with behaviour resembling that of a wire—to an 'active integrator' [276]. Through this *active* role of neurons, Mukherjee (p. 282-3 [276]) explains that we are now able to 'imagine building extraordinarily complex circuits… the basis for… even more complex computational modules those that can support memory, sentience, feeling, thought, and sensation… [and that] could coalesce to form the human brain.' Such language brings computational studies to mind, and indeed, the dawning awareness that networks of astrocytes had the potential to contribute to long range signaling around 1990, supplemented by experimental evidence over the subsequent decades (see [277, 278] and references within), has seen the advent of a field dedicated to computational investigations of the interaction between glial matter and neurons [279].

**15.2. Brain development and the growth of consciousness**

The growth of consciousness from a single cell to the highly differentiated brain of a complex organism on a timescale of months to years, prompts questions like: What creates the incredibly intricate complex of neural cells that support the almost magical experience of consciousness? And when can consciousness be claimed to have arisen [280] or been lost? This latter question is laden with ethical consequences (in the context of the termination of life support, for example), begging consideration of how the presence of consciousness can be identified experimentally (see, e.g. [281] for a relevant discussion). For patients unable to respond directly, including through eye movements, conscious activity must be proxied by other means, notably the measurement of activity in the brain.

From a psychological perspective, the development of consciousness in humans is a process that takes place over time, and that is associated with specific landmarks—such as the attainment of a sense of self. These landmarks necessarily correlate with physical changes taking place within the brain.

With respect to our consideration of the role of white matter, we note that myelination commences *in utero*, continues through childhood, and that the period of maximum myelin growth coincides with the development of executive functions in late adolescence and early adulthood. This latter period coincides with the onset of many psychological disorders associated with abnormal white matter development. Already in the abstract [264], Haroutunian points out the significance of the abundance of white matter in the human brain and the role that development plays in mental health: '… we highlight the role of glia, especially the most recently evolved oligodendrocytes and the myelin they produce, in achieving and maintaining optimal brain function.' He clarifies: 'The human brain undergoes exceptionally protracted and pervasive myelination (even throughout its GM) and can thus achieve and maintain the rapid conduction and synchronous timing of neural networks on which optimal function depends. The continuum of increasing myelin vulnerability resulting from the human brain's protracted myelination underlies underappreciated communalities between different disease phenotypes ranging from developmental ones such as schizophrenia (SZ) and bipolar disorder (BD) to degenerative ones such as Alzheimer's disease (AD).' [264]

Limitations of space necessitate a highly selective discussion; we restrict ourselves to a consideration of the earliest part of the developmental period, relating the consequences of early birth to brain development, touching on the existence of consciousness at this time (see [282] and references therein). Childhood, a period of extensive learning associated with, for example, continued massive pruning of the synapses, is skimmed over, as is the growth of executive functions and brain maturation during the period to adulthood. Learning and experience, and emotional and psychological development are beyond the scope of this contribution. We are also obliged to ignore all that can be learnt about the diminishing sense of self and of conscious awareness attributable in varying degrees to progressive dysregulation of myelination from dementia, alzheimer's disease and other degenerative disorders.

*15.2.1. Infancy*

Although, some mammals undergo considerably longer pregnancies than humans (such as 645 and 590 days for the African elephant and the sperm whale, respectively), we stand out in the animal kingdom by only attaining adulthood a full decade after reaching reproductive maturity during adolescence. Having compromised on a nine-month pregnancy, evolution has ensured that term-born newborns are equipped with what they need to survive in the outside world: food-seeking behaviour, a sucking reflex, an ability to recognise their mother's voice from the sounds heard in the womb and, vitally, the ability to secure the attention and care of their mothers (from birth and for decades to come).

*15.2.2. Pre-term infants*

Deprived of the luxury of developing within the protective environment of the womb, babies born preterm emerge before they are ready to take on the challenges of the outside world. From the instant of birth, all of their senses are subjected to an unfamiliar, hostile environment. The preterm baby is deprived of a steady source of warmth and the familiar taste of amniotic fluid, and no longer benefits from the filtering that gives rise to a suffused pink glow and muffled sounds. These infants experience a harsher environment: cool air on damp skin, unfiltered light on thin eyelids, the loud noises of the delivery room, and the unfamiliar sensation of being handled. They become overstimulated easily. One example: they can focus on a black and white pattern held close to them, but unlike a term-born child, they cannot break their gaze by looking away once their attention is saturated [283]. Thus the infants' immature minds encounter significant challenges at a time when their bodies are having to cope with life outside the womb.

Historically, the outcome for those significantly premature babies that survived has been relatively poor, and not only as their due date passed, but even years afterwards: their physical development and academic achievements lagged on those of their term-born peers.

Prior to birth, essentially drugged by their environment, foetuses spend most of their time asleep [284]. Foetal sleep patterns develop and sleep changes during pregnancy, for example, the characteristic loss of muscle tone associated with rapid eye movement, or REM, sleep does not appear until late in pregnancy (see, e.g. [285] and [286]). By 23 weeks a foetus will spend roughly 6 hours in REM sleep, 6 hours in non-REM sleep and the remaining 12 hours in an interim sleep form. Only in the final weeks prior to birth do babies spend any *significant* time awake, and even then, they sleep for all but two or three hours each day [284]. The relative proportion of REM sleep increases during pregnancy too, until a couple of weeks prior to term the foetus engages in some nine hours of REM sleep per day. This increases to a full twelve hours in the final week, which is more REM sleep than observed at any other period in life. It is also a time of massive synaptogenesis. As Walker emphasises, it is difficult to exaggerate the importance of sleep, and the extent to which sleep plays an active and vital role in our health, development and wellbeing in so many ways throughout life [284].

DiGregorio informs us that [283] 'The brain development that makes us uniquely human is accomplished in the last part of pregnancy. For premature babies, it must be accomplished in the NICU. Between 28 weeks and term, the fetal or premature brain triples in weight.' Amongst other significant findings, the Developing Human Connectome Project has revealed that: 'The early developmental disruption imposed by preterm birth is associated with extensive alterations in functional connectivity.' [287] Lagercrantz expresses the opinion that infants have minimal consciousness at birth, but also that 'even the very preterm infant may be more conscious than the fetus of corresponding gestational age' [288]. Thus, the development of consciousness will be significantly affected by premature birth (see, e.g. [282] and references therein).

The implementation of 'kangaroo care', whereby premature newborn infants are held in direct contact with the skin of a parent or carer in a quiet, darkened room—essentially attempting to reproduce the environment of the womb—has improved outcome, including over the long-term [280, 289]. The environment is only part of the story, however, and it is known that genetic factors, birthweight and adversity can affect telomere length (the length of the protective ends of linear chromosomes). Low birthweight newborns have a *shorter* mean telomere length than typical newborns of the same gestational age [290]; the telomere length is *longer* at birth, but decreases *disproportionately rapidly* for premature infants compared to term-born ones [291–293]. A shorter telomere length indicates a heightened risk of developing dementia, certain types of cancer, and cardiovascular and metabolic disorders, including chronic hypertension and hyperglycemia. Okuda notes that 'variations in telomere length among adults are in large part attributed to determinants (genetic and environmental) that start exerting their effect in utero' [294]. With respect to the urgency of returning children to the classroom during the pandemic—and enabling adults to resume a more normal existence—it should be noted that telomere length is diminished by living under extremely adverse conditions, including needing an intensive care unit, and living in an orphanage (e.g. [292, 293, 295]).

*15.2.3. Term-born infants*
In their first weeks of life, infants focus best at a distance equivalent to that of the face of their mother while nursing, enabling them to react to faces, and to mirror movements. Newborns respond to the primary features of a face—two eyes and a mouth, irrespective of whether the features are presented upside down or the 'right way' up. Infancy and early childhood are key developmental ages in multiple respects, with psychological development running in parallel with motor development, language aquisition and learning in general (see Figure 11 [250, 296, 297] and references therein). With such rapid development, the early months and years offer rich evidence of how the human mind is formed [253], but they also present exceptional experimental challenges as infants, toddlers and young children are unwilling to be constrained and too young to reason with successfully (see, for example, the attrition rate in [298]). The acquisition of MRI-based brain scans requires subjects to remain still, with their heads in a fixed position in a noisy, unfamiliar and claustrophobic environment. Microstructural white matter development, for example, is investigated using diffusion MRI, which requires that subjects remain static throughout the lengthy scanning process which is problematic for investigations of young children. This experimental challenge has meant that despite the evident interest in understanding the structural changes that take place early in life, it is a period for which relatively few comprehensive studies of the brain exist. Despite the experimental obstacles, tremendous advances are now being made, see e.g. [299, 300]. Furthermore, technical advances such as the development of caps and portable wireless headsets for toddlers and young children facilitating unconstrained movement while data is collected, are, for example, enabling electroencephalographic studies in extremely preterm infants [301].

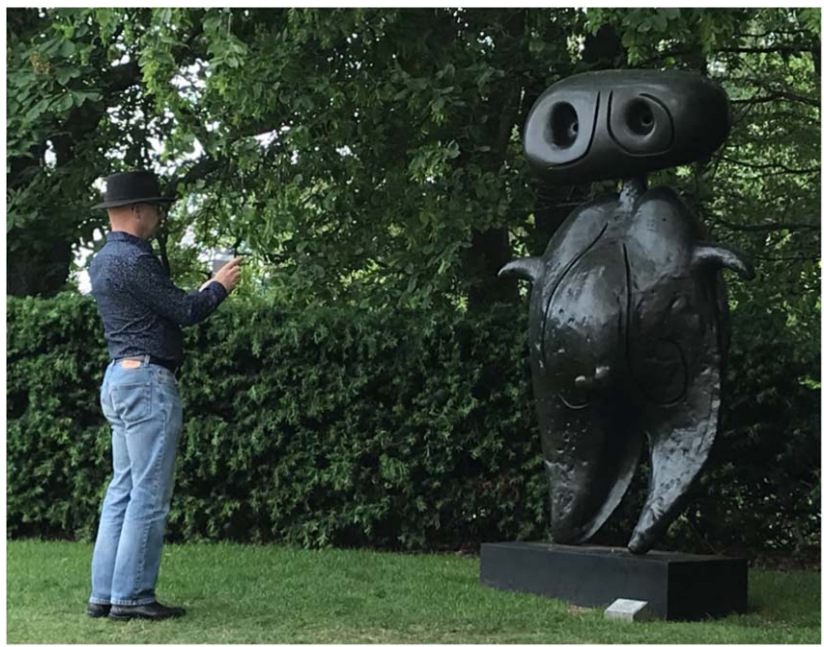

**Figure 11.** Joan Miró's explorations of humankind's true identity resulted in a sculpture, *Personnage*. In this photograph, taken at Louisiana Museum of Modern Art in Denmark, *Personnage* has caught the attention of Brain Prize recipient Stanislas Dehaene. Dehaene is the author of several books and papers on consciousness, including *Consciousness and the Brain* [250]. Photo: Suzy Lidström, 2019.

Researchers have managed to overcome extreme experimental challenges to probe aspects of cognition in infants successfully, assessing the extent to which brain activity is similar to or differs from that of adults. Specifically, for instance, infants aged 2 to 9 months are able to utilize the specialised neurons [302, 303] that adults employ to recognise faces [260], bodies and places (see [304] and references therein) albeit with important differences [305]. Within months of their birth, babies exhibit similar preferences to those of adults for large-scale organization of key categories in the visual cortex. Presented visually, these categories evoke brain activity 'within circumscribed, highly selective regions' as well as producing 'graded response patterns across larger swaths of cortex', but with the areas involved being less specialised than those in the adult brain.

The determination of whether responses are learnt or innate will advance understanding of the developmental progression from imitation to identification, a key developmental step on the path to attaining a sense of self. Magnetoencephalography has revealed that, by seven months, babies are already able to distinguish between themselves and others to some extent, as evidenced by their reaction when being touched themselves and when watching another person being touched [306].

And what is known of the emergence of consciousness in the first stages of life? Lagercranz and Changeux find it difficult to believe that the foetus is conscious to any considerable degree, despite exhibiting specific markers of consciousness, like a limited degree of self awareness [288], and for them, the absence of 'fully estabished' thalamocortical connections determines that the newborn is only capable of attaining a minimal level of consciousness [280] .

**15.3. Development**

The brain of a newborn contains less than one third of the synaptic connections that an adult has. Furthermore, 'The connections it does have are mostly eliminated and replaced in the first year of life' (see Sam Wang, pp 34-59 in [252]). Specific windows of opportunity open during development, enabling different capabilities to be mastered successively. Binocular vision is one such [307, 308], language another. In infancy, children are capable not only of mastering their mother tongue, but also other languages to which they are exposed. This ability is lost by adolescence. Interestingly, though, children who have not grown up to master a language to which they were exposed as an infant (for example because they were subsequently adopted 'abroad') will still respond to linguistic elements from the language to which they were initially exposed in a manner that indicates that the now unfamiliar language is being processed as a language and not just as sounds [309]. There is also window for the development of empathy and emotional behaviour: After the fall of communism, many severely deprived children in Romanian orphanages were adopted [310]. Prior to adoption, they had received perfunctory care and experienced a complete deficit of love and empathy, as well as inadequate sensory stimulation. As time passed, it became apparent that, while the youngest infants and children might rally, showing a typical developmental trajectory upon adoption, children adopted beyond the age of four were locked into a mode of behaviour reminiscent of that of a child on the autistic spectrum. To understand why this comes about, the process of normal brain development needs to be understood. Interestingly, recent research on empathy reveals how activities and interactions in play provide an insight into who children become [311].

The impact of learning, and education, experience and motivation and the associated physical changes in the brain are beyond the scope of this piece. Suffice it to say that we are born into our respective environments with a low level of consciousness, but with a brain that has a massive capability to develop in response to our experiences, interactions and passions. Initial massive synaptogenesis is replaced by widespread pruning, as we shape our own brains. As we develop, we pass the milestones identified by psychologists until we reach adulthood. In doing so, a set of regions closely associated with the very processes that we feel make us who we are exhibits *depressed activity* when an adult is involved in goal-oriented tasks. This 'default network', is believed to support mental activity that is introspective, self-referential, stimulus independent, self-projecting, etc [312]. The architecture of this network both strengthens and becomes more well developed from late childhood into early adulthood, at a time when, unsurprisingly, introspective and stimulus-independent thought is developed most strongly [312].

### 15.4. Consciousness and physics

Understanding consciousness is one of the greatest challenges of our time. In the context of mathematics, having asked what breathes the fire into equations, Stephen Hawking pointed out:

> The usual approach of science of constructing a mathematical model cannot answer the questions of why there should be a universe for the model to describe. Why does the Universe go to all the bother of existing?
>
> *Stephen W Hawking* [8]

The aforementioned stages in the development of a human brain represent the highest known level of the development of consciousness across the spectrum of animals down to birds and even insects, depending on one's definition of consciousness. It is absurd to say that all of Nature is conscious, even down to the level of an inorganic mineral, but consciousness surely grows out of related natural phenomena (see [253] for a philosophical perspective). If the ordinary inanimate phenomena of nature are analogous to the incoherent oxidation of inorganic matter, consciousness is analogous to the coherent blazing conflagration of a widespread forest fire.

Although the 'perennial appeal of quantum approaches to consciousness' is dismissed in 'Putting Descartes before the horse' [313] and in Tegmark's more comprehensive consideration of decoherence time [314], these authors have noted that there might indeed be a need for new physics. Seth explained, 'the means by which neurochemical brain activity engenders subjective conscious experience… can still seem entirely mysterious, and perhaps requiring scientific revolution rather than evolution'.

The adoption of an interdisciplinary, holistic perspective will be required to comprehend the integrated operation of white and grey matter, as promoted by Haroutunian [264]. In respect of such an approach, the Director of the American National Institute of Health noted:

> … new advances in computer science, math, and imaging and data visualization are empowering us to study the human brain as an entire organ, and at a level of detail not previously imagined possible in a living person.
>
> Some have likened this new ability to the difference between listening to the string section (evaluating an isolated part of the brain) versus listening to an entire orchestra (the whole organ). If you listen only to the string or percussion section, you'll gain a pretty good understanding of how that particular group of instruments sounds. However, that experience would not compare to the experience of listening to the whole orchestra and chorus perform Beethoven's Symphony No. 9, the Ode to Joy.
>
> *Francis Collins in 'The Symphony Inside Your Brain'* [315]

We postulate that, if the orchestra represents the brain and its operations, then the appropriate analogy for consciousness is the music as it is carried through the air.[26]

Our title recasts, in the context of consciousness, Stephen Hawking's reflection on how mathematical physics is able to encapsulate vibrant physical concepts:

---

[26] Roland Allen, private communication.

> Even if there is only one possible unified theory, it is just a set of rules and equations. What is it that breathes fire into the equations and makes a universe for them to describe?
>
> *Stephen W. Hawking* [8]

Our response is that just as Nature breathed fire into the equations, it is Nature that breathes the fire of consciousness into our brains, a response that reflects the scientists' quest, as exemplified in a 2021 Nova broadcast on entanglement, when Anton Zeilinger, a coauthor on this paper, remarked: 'I am just trying to understand Nature'.

## 16. What philosophers should *really* be thinking about by Roland Allen and Suzy Lidström

Philosophy over the centuries has changed its meaning, and philosophy in the twenty-first century essentially means the struggle for clear thinking. It is obvious that the profound difficulties of human society have been and continue to be consequences of misunderstandings and even delusions, so—as is already widely recognised—philosophers can provide guidance for a better future by pointing out the common errors in practical matters, ranging from grossly harmful cultural and religious practices to the more subtle nuances of ethical behaviour. Here, we would like to extend this enterprise to the highest imaginable levels in trying to interpret what is correct and erroneous in the scientific enterprise and what are the deepest underlying principles of this enterprise.

Our first example is the separate concepts of a multiverse and the anthropic principle, which it is now fashionable to either accept or abhor for what are usually the wrong reasons. For example, in what is otherwise still the best broad treatise on cosmology [316], a footnote states

> It is unclear to one of the authors how a concept as lame as the 'anthropic idea' was ever elevated to the status of a principle.

This is a psychological and even emotional rather than scientific statement. In the other direction, the concept of landscape has come to be regarded as a positive feature of, and even justification for, string theory, but other theories may also have landscapes and the landscape may not be necessary to understand why our universe seems favourable for the development of intelligent life [317]. Furthermore, string theory itself has developed largely because its mathematical beauty is appealing to human theorists. (The beauty of a theory is not directly relevant to its physical correctness.) So what are the valid arguments that would lead to accepting the potential reality of a multiverse? The first would be experimental verification. For example, in the Everett interpretation of quantum mechanics—which implies one type of multiverse—demonstration of entanglement in many different contexts and with higher and higher levels of macroscopic physical systems would imply that the standard quantum formulation does apply to all physical systems including human observers. If the multiverse is an unavoidable logical implication of an accepted physical theory then acceptance of that theory inescapably implies an acceptance of the multiverse. In the vast number of papers that have been written on the interpretation of quantum mechanics it is evident that nearly all betray lapses in clear thinking on some level. So, to review, philosophy can help the scientific enterprise, by emphasising what are valid and invalid logical arguments.[27] It is invalid to rule out a multiverse, or any physical theory because one dislikes it. It is invalid to cherish a physical theory because one finds it intellectually appealing. It is valid to accept a physical theory if it is confirmed by experiment or if it is inevitably implied by accepted theory. There are, of course, many other examples besides the multiverse of how these principles apply and should be routinely employed in science.

In addition to reminding scientists of what is invalid and valid reasoning, philosophers may look for principles that are too deep to be considered in normal scientific thinking and publications. One such principle is this: How do we understand what systems or concepts emerge to be dominant in Nature or human society? For example if there is a multiverse with $10^N$ potential universes, with N greater than 1000, how do we estimate that such a universe will be stable at all and will be sufficiently stable to harbour intelligent life? If we assume the best version of Nature currently available, namely the path integral description of either quantum field theory or a deeper theory, then a solution is provided by the power of the exponential function in either the Lorentzian formulation, with large actions killed off by rapid oscillations, or the Euclidian formulation, with large actions killed off by exponential decrease. Even an extremely large number of unstable universes lose in probability to those that are stable.

---

[27] We consider Richard Feynman and Murray Gell-Mann to be the greatest theoretical physicists of their generation. Feynman is on record as saying with regret that he initially disliked the gauge theory of the standard model because he regarded its asymmetry as ugly [318]. One of us speaking to MGM mentioned that the primal feelings of human consciousness are not fully explained by the current mathematical laws of physics to which Gell-Mann responded 'That's crazy talk' [319].

One can imagine extending this principle to weighting the factor involving the action by a similar factor which expresses the probability for the development of intelligent life. In a multiverse scenario this provides a quite respectable foundation for even the anthropic principle. In this context it should be mentioned that Stephen Weinberg predicted the approximate value for the dark energy [93] before it was discovered [76, 320].

This general idea, of the stability of an occurrence dominating the sheer number of occurrences also explains many other observations in Nature and human society. Why is a single kind of molecule, DNA, the basis of all life on Earth, whereas an infinite number of molecules can be formed from the common elements? Because DNA has been proved to be stable on a timescale of billions of years. Why is the Riemann hypothesis still regarded as the greatest problem in mathematics [4], as it was in 1900 [321] and again in 2000 [322]? Because the worldwide community of mathematicians is justifiably in awe of the profound connections between number theory and the deepest other aspects of mathematics.

Our chief points are: (1) As scientists who have observed at close hand and in publications that even the best scientists often display remarkable philosophical naïvete in their reasoning, we believe that some interaction between philosophers who understand science and the scientific community could have a true positive impact. (2) Another potential role for philosophers is in considering the deepest principles behind science—such as those given immediately above—in clarifying these issues, Richard Feynman was profoundly sceptical of philosophers, thinking that philosophers would provide vacuous explanations for scientific facts, possibly thinking of Hagel and his apparent belief that philosophical arguments can limit the number of planets. But 21$^{st}$ century philosophers with greater sophistication and an understanding of the real fundamental principles in science can offer some positive influence in scientific thinking.

## 17. How can scientists address misinformation? Science, survival and the urgent pursuit of truth by Steven Goldfarb

We live in challenging times. At the time of writing, the world is attempting to navigate its way through rapid climate change [323], a global pandemic [324] and economic collapse [325], all in the backdrop of increasing socioeconomic inequality [326] and depleting resources [327]. It is at times like this that we turn to scientists for solutions, and to our world leaders to provide resources and guide the implementation of these solutions.

Indeed, international teams of scientists have heeded the call, coordinating their efforts to find solutions that are both effective and safe, then communicating them to the world leaders. Researchers in climate science, epidemiology and economics have all spoken up, noting the importance and urgency of the problems at hand, and offering paths forward. Those of us in fundamental research support these efforts, through public talks and editorials, and sometimes by advocating to politicians and other key stakeholders.

Despite these efforts, the scientific advice is not always heard. Worse, even when it is heard, it is often ignored. Why? Although most of the world's leaders have realised the urgency of the situation and have used their skills at communication and consensus-building to motivate their citizens to work together for the common good. Others clearly have not.

By not taking action or by taking inappropriate action, these leaders are endangering their own constituents, future generations, and quite possibly an entire species. As scientists, we have a moral obligation to expose these misdeeds and to inform the population of the actions that need to be taken. Unfortunately, this is not easily done.

Toward the end of the last century, scientists working at CERN, the international particle physics laboratory in Geneva, Switzerland, developed a communication application designed to facilitate the sharing of scientific documentation around the globe [328]. This tool, the World-Wide Web, has more than served that purpose, allowing instant sharing of information not only between science institutions and laboratories, but between individuals everywhere. As the reader is well aware, a wealth of information, knowledge and wisdom is now available, quickly and affordably, to nearly everyone on this planet.

Ironically, it is this very tool that is at the heart of the problem. Belligerent and/or ignorant parties are able to use social media platforms on the web to disseminate false or misleading information rapidly, while posing as reliable sources. Those with a natural penchant for communication, and minimal training, are able to exploit naïve, ill-informed or simply trusting audiences on these platforms, whether to push a political or financial agenda or simply to wreak chaos for the fun of it.

Anyone with an understanding of science or history knows that, in most cases, facts and evidence eventually do come to light. However, a lot of damage can occur in the meantime. Thus, many scientists are taking pro-active approaches to address the issue. Some have honed their communication skills, interacting through the traditional media, while others (often younger researchers) develop social media strategies to effectively disseminate scientific advancements and knowledge.

Surprisingly, as a particle physicist, I find myself dedicating nearly as much time to current political discourse as I do to current research. There is a lot at stake and we, as scientists—people who have dedicated our lives to the pursuit of truth—cannot afford to ignore it. At the heart of the issue is the human ability to discern truth from fiction. Those with well-amplified, far-reaching communication platforms have the ability to disguise lies as truth and vice-verse, confusing audiences and undermining public trust in science. This is a hard battle to fight.

In one of my more recent presentations [329], I spent a significant time describing the complex and rigid process researchers follow to go from basic idea to publications. My hope is to instil an appreciation for science that can transcend the disinformation that bombards us every day, by explaining the effort required to attain truth, and thus the value of science to humanity.

Such efforts can have an important effect in the short term, but in many cases there are simply too few science communicators or resources to battle with professional liars. It is much easier to spout untruths at random, than to do research and present the results to the public in a convincing manner. This problem is compounded when the sources of the misleading information are in positions of power or are members of the professional media.

Fortunately, our nature provides a path for a long-term, sustainable solution. Human beings have a natural affinity for science and discovery, especially at a young age. Our DNA provides us with the means to address certain basic needs: finding food, building shelter, making babies, and seeking a better understanding of our universe. It is the last capability that allows us to develop and improve the tools needed by future generations to survive.

This instinct motivates us to create art, music and literature, and pursue science. It is driven by our inherent curiosity, but goes deeper than that, in that it compels us to share our findings with our family, friends and fellow inhabitants. That is, we are all scientists from the day we are born. As discussed in an earlier contribution to this paper, as soon as our eyes open, we look around, take in our environment and try to make sense of it.

Our current environment, however, does not always provide equally fertile ground to develop this ability. What varies from person to person is our understanding of the existing knowledge base, the proposed models to describe it and make predictions, and the methodologies employed to build these models from the data. The knowledge and skill sets we attain depend on individual capability, experience and access to quality education. Thus, there is an important socio-economic aspect we cannot ignore.

As young children, we are fascinated by the beautiful blue sky. We share that fascination with those around us, who confirm that they also see a blue sky, and teach us the name of the colour. Before long, we wonder why the sky is blue. Then, if we are fortunate, after years of formal education, we might learn the formalisations needed to understand the transmission and scattering of sunlight [330], optics, electromagnetism and waves.

Unfortunately, somewhere along the way, between kindergarten and elementary electromagnetism courses, many lose the thread connecting the initial thrill of discovery to the formal education required to develop a deep understanding of its meaning. Great teachers recognise this and do what they can to bring that thrill back to the classroom. Some are able to relate a lesson to their own research, or to current science headlines. But this is not always an easy task, and often the latest headlines involve seemingly complex topics unfamiliar to the teacher or the students. Furthermore, many teachers do not have access to that information.

This is where informal science education can make an impact. Much modern research is anchored in basic concepts. An appreciation of this enables those who are active in public engagement to convey the fundamental aspects of recent advancements in language that is accessible to the general public: Dark Matter and conservation of momentum, the Higgs boson and a cocktail party, gravitational waves and billiard balls on a sheet, viral infection and dominoes. By working together with formal educators, these scientists can bring the excitement of current research to the classroom and use these concepts to catalyse the learning process.

Although the current reach of informal science programs is still rather limited, they are growing in size, scope and worldwide reach. The International Particle Physics Outreach Group (IPPOG) [331], for example, runs the International Particle Physics Masterclass and Global Cosmics programs reaching tens of thousands of students in 60 countries around the world. These programs partner active researchers with secondary school teachers to give their students the possibility to learn what it is like to be a scientist today.

The students become actively involved in the research, analysing actual data from current particle physics or astrophysics experiments. This has the effect of re-igniting that flame of curiosity from childhood or, in some cases, fanning existing flames sufficiently to spark interest for future studies. Most importantly, students learn the methodology employed by scientists to explore data and to address the complex problems they are trying to solve. That is, they re-learn the scientific process and the value of evidence-based decision making.

This is no small step. Students exposed to these opportunities develop an appreciation for science and the scientific process. Through improved understanding of the thought processes they, as individual citizens, are better prepared to sift through the mountains of lies they are presented each day, to find the facts they need to fuel their decisions. As they mature, they will be better able to choose appropriate sources of information, and will demand the replacement of deceptive leaders by people keen to govern based on evidence and valid argumentation.

It would be an exaggeration to think that such efforts will have dramatic effects in the short-term. Trolls are certainly here to stay. There will always be people who feel sufficiently disenfranchised to want to break existing power structures through lies and deception. Only significant global economic and political change can address the underlying issues there. However, the effect of their weapons can be greatly reduced in the long-run through education and improving the fundamental understanding of science by future generations. And this battle is being fought today.

## 18. Can we find violations of Bell locality in macroscopic systems? by Bryan Dalton

To Einstein [51], the Copenhagen quantum interpretation of what happens when we first measure an observable $\Omega_A$ in one sub-system *A* with outcome $\alpha$, and then immediately measure an observable $\Omega_B$ in a second well-separated sub-system *B* with outcome $\beta$ seemed counter-intuitive, implying 'instantaneous action at a distance' during the two-step measurement process. This has been known since the 1930s as the EPR paradox. According to the Copenhagen interpretation, after the first measurement, the quantum state is changed, conditioned on the outcome of the first measurement. As a result, the reduced density operator describing the original state for sub-system *B* would have changed instantaneously to a different state, despite no time having passed in which a signal could have travelled between the two well-separated sub-systems. This effect is referred to as steering [332]. Of course if $\Omega_A$ was immediately measured a second time, it is easy to show that the outcome $\alpha$ would occur with probability 1. For the Copenhagenist, this raises no issues, since the quantum state is not regarded as a real object, but only a means of determining the probabilities of the outcomes of measuring observables (the outcomes being the real objects which are created by the measurement process on the prepared quantum state). That the quantum state changes as a result of the measurement of $\Omega_A$ with outcome $\alpha$, merely signifies the probability changing from its previous value for the original preparation process, to now being unity for a new preparation process in which the second part involves measuring $\Omega_A$ with outcome $\alpha$. If we now measure the second sub-system observable $\Omega_B$ the conditional probability for outcome $\beta$, given that measurement of $\Omega_A$ in the first sub-system *A* resulted in outcome $\alpha$, will now be determined from the new conditioned quantum state. In general this will be different from the probability of outcome $\beta$ resulting from measurement of observable $\Omega_B$ for the original quantum state. However, using Bayes' theorem the joint probability for outcomes $\alpha$ for $\Omega_A$ and $\beta$ for $\Omega_B$ can be determined to be the standard Copenhagen expression for the joint measurement probability for the measurement of the two observables in the separated sub-systems if the measurements had been made on the original quantum state totally independently of each other and in no particular order. As far as we know, the predictions based on the Copenhagen version of quantum theory are always in accord with experiment. But to Einstein and others, the Copenhagen theoretical picture was philosophically unsatisfactory. The question arose: is it really necessary to invoke the Copenhagen picture involving the instantaneous change to the quantum state as a result of the first measurement (the 'collapse of the wave function') to describe what happens, or is there a simpler picture based on classical probability theory—and involving what we now refer to as hidden variables—that could also account for all the probability predictions of quantum theory?

A first question is whether the results for any quantum states describing two sub-systems can also be described by hidden variable theory. One whole class of quantum states that can be so-described are the separable states [333]. Here the initial process involves preparing each separate sub-system in a range of sub-system quantum states, each choice being specified according to its probability. However, the results for the joint measurement probability outcomes for $\Omega_A$, $\Omega_B$ are of the same form as in local hidden variable theory. So as the separable states can all be given a local hidden variable theory interpretation, it follows that any state that cannot be so interpreted must be a non-separable or entangled state. However, Werner [333] showed that there were some entangled states that could be interpreted in terms of local hidden variable theory. Particular examples were the so-called Werner states [333], which are mixed states specified by a single parameter, involving two sub-systems with equal dimensionality. This means that the division of quantum states into separable or entangled ones does not coincide with their division into Bell local and Bell non-local. The separable states are examples of quantum states that can be also described by local hidden variable theory, and are characterised by both sub-systems being associated with a so-called local hidden quantum state [334] which is specified by the hidden variables $\lambda$. Clearly within local hidden variable theory we could also have the situation where only one of the two sub-systems, *B* say, is associated with a local hidden quantum state from which the measurement probability for the outcome for $\Omega_B$ is determined; for the other sub-system, *A*, the corresponding probability for outcome $\alpha$ for $\Omega_A$ is not determined from a local hidden state. Another situation is where neither sub-system is associated with a local hidden quantum state. Both of the latter situations involve entangled quantum states, whilst still being described by local hidden variable theory. States where there are no local hidden states are referred to as EPR steerable states [334]. They allow for the possibility of choosing the measurement for observable $\Omega_A$ to steer sub-system $B$ such that the outcome for measuring $\Omega_B$ can be chosen in advance. The EPR steerable states are all entangled, and include those that are Bell non-local as well as some that are Bell local and entangled (named after John Bell, shown in figure 12), and are said to exhibit EPR correlations. Bell non-locality is the most general form of hidden variable theory for describing the two sub-systems. Here there are no separate hidden variable dependent probabilities for sub-system observable measurements. To determine whether a state is Bell non-local it must be shown that a Bell inequality—derived from the basic hidden variable expression for the joint probability—is violated.

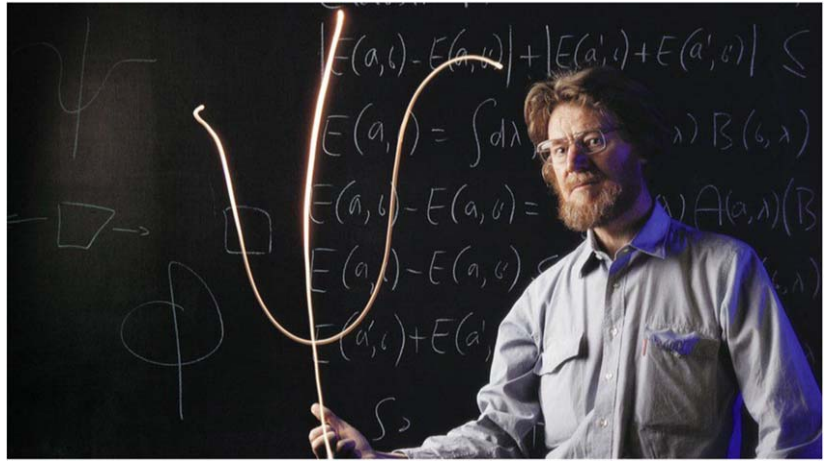

**Figure 12.** Based on a local hidden variable theory interpretation of quantum theory, John Bell (pictured here at CERN in 1982) derived a famous inequality involving correlation functions for the measurement outcomes of spin component measurements on two spin 1/2 sub-subsystems, that could be tested against the predictions of quantum theory in the case of an entangled singlet state of the combined system. This inequality was shown theoretically to be violated for certain spin component choices, and this result was later confirmed in experiments. Credit: CERN PhotoLab.

As pointed out recently [335], there are a multitude of Bell inequalities that can be derived even for bipartite systems, depending on the number of observables considered in each of the two sub-systems and on the number of different outcomes for each observable. One of the earliest of these was the famous CHSH Bell inequality [336]. Here there were two different observables $\Omega_{A1}, \Omega_{A2}$ and $\Omega_{B1}, \Omega_{B2}$ for each sub-system, and measurement of any observable was restricted to two outcomes—which we choose to be $+1/2$ and $-1/2$. The CHSH inequality is $|S| \leqslant 1/2$, where $S = \langle \Omega_{A1} \otimes \Omega_{B1} \rangle + \langle \Omega_{A1} \otimes \Omega_{B2} \rangle + \langle \Omega_{A2} \otimes \Omega_{B1} \rangle - \langle \Omega_{A2} \otimes \Omega_{B2} \rangle$. Suitable physical systems for which this inequality can be tested include two spin 1/2 sub-systems, with components of the spins along various directions being the observables since the measured outcome is either $+1/2$ or $-1/2$. Another suitable physical system is two modes of the EM field as the two sub-systems are each occupied by one photon, with the mode polarisation being the observable and the outcome being $+1/2$ or $-1/2$ depending on whether the outcome is right or left circular polarisation, or up or across for linear polarisation. These examples are both microscopic systems. Experiments testing the CHSH inequality have been carried out since the 1970s (see [335] for a recent review), and a violation of the inequality has now been convincingly demonstrated following numerous improvements to remove possible loopholes by means of which the inequality might not really be violated.

However, apart from situations involving two super-conducting qubits, the CHSH inequality only establishes Bell non-locality in microscopic systems. As quantum theory was originally formulated to treat microscopic systems, merely showing that the Copenhagen interpretation was needed for microscopic systems leaves open the possibility that hidden variable theory could still be used to explain experimental effects in macroscopic systems. The latter, after all, normally lie in the domain of classical physics where quantum theory is not usually required. Hence there is an interest in finding quantum systems on a macroscopic scale for which Bell inequalities can be derived, and for which violations might be both predicted and found experimentally. There are examples from the 1980s of Bell inequalities applied to macroscopic systems, though no experimental tests have yet been carried out. In [337] a system consisting of two large spin $s$ sub-systems was considered allowing for measurements of any spin component to have outcomes from $-s$ to $+s$ in integer steps. For an overall singlet pure state in which measurement of a spin component in one sub-system leads to the opposite outcome when the same spin component was measured in the other, a Bell inequality involving spin components along three unit vectors $a,b,c$ of the form $s|\langle S_{Aa} \rangle - \langle S_{Bb} \rangle| \geqslant \langle S_{Aa} \otimes S_{Bc} \rangle + \langle S_{Ab} \otimes S_{Bc} \rangle$ was found. This was shown theoretically to be violated for coplanar unit vectors, where $a,b$ make an angle $\pi - 2\theta$ with each other and the same angle $\pi/2 + \theta$ with *c*, provided $0 < \sin\theta < 1/2s$. This is a very small range of violating angles if *s* is large enough for the system to be considered macroscopic, and the required singlet state would be difficult to create. In [338] two sub-systems each containing two bosonic modes were considered. An overall entangled state with a large number *N* of bosons was studied, and a Bell inequality found involving sub-system boson number-like observables for each sub-system. These were given by linear combinations (specified by a parameter $\theta$) of its pair of mode creation operators and raised to power *J*, times a similar expression involving the annihilation operators. For $J = N \to \infty$ the inequality is violated for finite $\theta$ if $3g(\theta) - g(3\theta) - 2 > 0$, where $g(\theta) = \exp(-J\theta^2/2)$. Although suitable $\theta$ can be found, the measurement of the observables for large $J = N$ would be difficult. Subsequently, Leggett and Garg [339] developed a test for macroscopic quantum coherence based on the mean value of products of pairs of observables for the two sub-systems, but now taken at three different times.

More recently, the interest in finding Bell non-locality in macroscopic systems has revived [340, 341]. This is in part due to experimental progress in the study of ultracold atomic gases, which are macroscopic systems for which a quantum description is required. These include ultracold bosonic gases, where large numbers of bosonic atoms may occupy each mode, creating Bose–Einstein condensates. For studying bipartite Bell non-locality, two mode systems such as those for bosons trapped in a double potential well, or for bosons in a single well but with two different spin states are available. A four mode bipartite system involving two modes associated with different internal states in each well can also be prepared [342] using atom-chip techniques. Multipartite systems in which each two-state atom is located at a differerent site on an optical lattice have also been created [343]. For ultracold fermionic gases the situation is not so clear, for although many fermion systems would be macroscopic, each mode could only be occupied by fermions with differing spins and hence many modes would be involved making it difficult to devise bipartite systems. Recent discussions of Bell non-locality in many-body systems are presented in [335, 344–346] and [347]. These contain examples of multipartite Bell inequalities, with applications to systems such as *N* two state atoms located at different sites in an optical lattice. Here each identical atom *i* is treated as a distinguishable two-mode pseudo-spin sub-system. Measurements on one of two chosen spin components $M_{i0}$ or $M_{i1}$ for the *i*th atom sub-system are considered, the two possible outcomes being designated as $\alpha_i = \pm 1$. Defining $S_0$, $S_{00}$, $S_{11}$ and $S_{01}$ involving the mean values of single measurements on individual spins or joint measurements on different spins, a Bell inequality $2S_0 + S_{01} + 2N + (S_{00} + S_{11})/2 \geqslant 0$ has been derived [344]. Bell correlations based on this inequality have been found [347] in systems involving 5 x $10^5$ bosonic atoms. In these systems the indistinguishability of the identical atoms and the effect of super-selection rules that rule out sub-system states with coherences between different boson numbers can be ignored, as there is just one atom in each separated spatial mode on each different lattice site. However, the symmetrisation principle and the super-selection rules are important in regard to tests for quantum entanglement and EPR steering [348, 349] in situations where the sub-systems must be defined via distinguishable modes rather than non-distinguishable atoms, and where there is multiple mode occupancy. The derivation of testable Bell inequalities for this common situation is an ongoing issue.

This section has been adapted from the more technical and more comprehensive treatment of this topic given in [350] (CC BY 4.0).

## 19. What is the source of quantum nonlocality? by Ana Maria Cetto

'That one body may act upon another at a distance through a vacuum without the mediation of anything else, by and through which their action and force may be conveyed from one another, is to me so great an absurdity that, I believe, no man who has in philosophic matters a competent faculty of thinking could ever fall into it.' wrote Isaac Newton, in a letter to Richard Bentley in 1692.

We live in a world full of 'signs' reaching us from the distance: The glitter of stars, the sound and strike of thunder and lightning, the pull of the Earth under our feet… Ancient deities, once endowed with supernatural powers to unleash such actions at a distance, have been left idle by the principle of locality, stating that objects are directly influenced only by their immediate surroundings. We have filled the void with a variety of fields surrounding the objects and mediating between them, to account for such actions no matter how distant the source–provided the speed of the message is not larger than the speed of light. Locality has been established as a basic tenet of physics

Or has it?

Quantum mechanics is conventionally said to have posed a challenge to locality. Bohm's rendering of the Schrödinger formalism is overtly nonlocal. Bell's theorem is widely interpreted as quantum mechanics outlawing local realism. (For other, not Bell-related inequalities that are violated by quantum mechanics, see [351] .) Experiments designed to test Bell-type inequalities with a pair of entangled particles (or photons) produce results consistent with the quantum predictions, suggesting the ruling out of local hidden-variable theories and consequently, of local realism altogether.

However, as is argued by a significantly increasing number of authors, this quite dramatic conclusion is based on a certain class of hidden variables pertaining only to the particle pair involved in the experiment. This is an unwarranted restriction, since, more generally, the variables could in addition describe a background field or medium that interacts with particles and detectors and intervenes in the measurement process. And, as demonstrated in [352], such 'background-based' theories can in principle reproduce the quantum correlations of Bell-type experiments.

An instance of a background-based theory is stochastic electrodynamics, the theory developed on the basis of the interaction of particles with the vacuum radiation field. The quantum features, as described by the Heisenberg and Schrödinger formalisms, emerge as a consequence of this permanent interaction [353] (for a comprehensive account of the first three decades of Stochastic Electrodynamics, see [354]). In a bipartite system, the particles become entangled by resonating to common field modes; the invisible, intangible vacuum acts as a mediator, and in turn becomes influenced by the particles [355, 356]. De Broglie's wave has an electromagnetic character: It is the modulated wave made up of the background waves at the Compton frequency in the particle's rest frame, with which the moving particle interacts resonantly (see [353] Ch. 9).

Is quantum mechanics the only instance in which the dynamics of particles is influenced by the surrounding medium, producing such 'nonclassical' behavior? Recent experiments with droplets bouncing on the surface of a vibrating liquid (see, e.g. [357, 358]) demonstrate that a background field can lead to a surprisingly wide range of quantum-like effects in the macroscopic, hydrodynamic realm, too. With each new bounce, the droplet contributes to form the pilot wave that moves along with it on the surface of the vibrating liquid.

In stochastic electrodynamics, as in the fluid mechanical quantum analogue, by hiding the underlying field element the description of the particle's behaviour becomes nonlocal. Of course, stochastic electrodynamics, although intriguing, has not been shown to replace either the qualitative or extremely precise quantitative predictions of quantum electrodynamics, but it reproduces them while providing a physically sound picture for the quantum formalism. And it explains the origin of the apparent nonlocalities.

## 20. How much of physics have we found so far? by Anton Zeilinger

The 20th century saw the discovery of two new big fields, the relativity theories and quantum mechanics. Could it be that similar, even larger discoveries are waiting around the corner? My, certainly not logically convincing, argument is the following. First we have to consider science in the modern way. In my eyes, it starts with the invention, if it is possible to say that, of the Renaissance point of view of the role of humankind in the Universe. During the Renaissance, humans started to dare to ask Nature questions. As I see it, a significant input was the rising self-esteem of humans as you see them in the gigantic change of portraits painted before and after the beginning of the Renaissance. Another interesting discovery was the discovery of Laws of Nature. Prior to the Renaissance, laws were God-given, and humans were not supposed to meddle in His work. Finally, we need to recall the great discovery that mathematics is the language of Nature. All these concepts led over the last few hundred years to immense discoveries, many new fields of science. We all are familiar with the development, step by step, of physics. But given that the development in science is of such a young nature compared to the history of humanity, I find it rather unlikely that we have discovered all the physics there is. I find it even unlikely that much of what we do know now will stand in the distant future. And, concluding, I hope that I am still alive when some young chaps discover the next great field.

## 21. Coda by Suzy Lidström

We conclude with the enduring voice of Stephen Hawking (see figure 13) as it was broadcast into space in a final message from the Cebreros antenna in Spain towards IA 0620-00, the closest black hole to Earth. Hawking's daughter, Lucy, described her father's message as being one 'of peace and hope, about unity and the need for us to live together in harmony on this planet'[28].

Hawking's message was directed at the young people whose task it will be to advance scientific frontiers and resolve the major challenges facing the world:

'I am very aware of the preciousness of time. Seize the moment. Act now. I have spent my life travelling across the Universe inside my mind. Through theoretical physics I have sought to answer some of the great questions but there are other challenges, other big questions which must be answered, and these will also need a new generation who are interested, engaged and with an understanding of science.

---

[28] http://esa.int/About_Us/Art_Culture_in_Space/ESA_honoured_to_take_part_in_Hawking_tribute

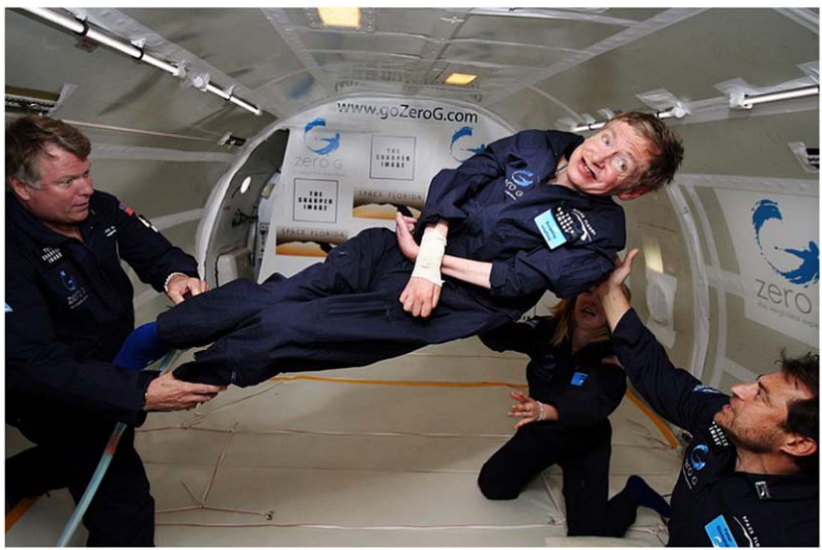

**Figure 13.** Stephen Hawking in zero gravity. Credit: NASA. Hawking's final words were broadcast into space set to music by the Greek composer Vangelis after his remains were laid to rest between those of Sir Isaac Newton and Sir Charles Darwin: 'We are all time travellers journeying together into the future. But let us work together to make that future a place we want to visit. Be brave, be determined, overcome the odds.'

How will we feed an ever-growing population, provide clean water, generate renewable energy, prevent and cure disease and slow down global climate change? I hope that science and technology will provide the answers to these questions, but it will take people, human beings with knowledge and understanding to implement the solution...

When we see the Earth from space we see ourselves as a whole; we see the unity and not the divisions. It is such a simple image, with a compelling message: one planet, one human race...

We must become global citizens...

It can be done. It can be done.'

After millennia of struggles in hundreds of cultures around the world to understand the Universe and our place in it, we are extremely fortunate to be living in a time when clarity is beginning to emerge. Our worldview is vastly grander than the narrow human-centred fantasies of past centuries. This article is meant to provide a microcosm of the best ideas that are surging through our current intellectual environment at the highest level. And as Hawking implies, with unparalleled eloquence, a central message is that equally grand challenges await even the youngest scientists who are just beginning to confront these issues today.

## Acknowledgments

The work of I K S, S F, O S, and A F was supported as part of the Quantum Materials for Energy Efficient Neuromorphic Computing (Q-MEEN-C) Energy Frontier Research Center (EFRC), funded by the U.S. Department of Energy, Office of Science, Basic Energy Sciences under Award # DE-SC0019273.

C F M, L A M C and P F thank Professor Lou Massa for his critical reading of their contrtibution to the manuscript and are grateful to the Natural Sciences and Engineering Research Council of Canada (NSERC), the Canada Foundation for Innovation (CFI), Mount Saint Vincent University, Université Laval, and Saint Mary's University for financial and material support. A A C would also like to acknowledge the financial support provided by NSERC of Canada

I K S, S F, O S and A F acknowledge the support received as part of the Quantum Materials for Energy Efficient Neuromorphic Computing (Q-MEEN-C) Energy Frontier Research Center (EFRC), funded by the U S Department of Energy, Office of Science, Basic Energy Sciences under Award # DE-SC0019273.

## Data availability statement

This is a perspective article that does not present novel data. The data that support the findings of this study are available upon reasonable request from the authors.

## ORCID iDs

Roland Allen https://orcid.org/0000-0001-5951-4717
Ana María Cetto https://orcid.org/0000-0001-6006-1102
Alan A Coley https://orcid.org/0000-0001-5482-6703
Bryan J Dalton https://orcid.org/0000-0002-1176-7528
Peyman Fahimi https://orcid.org/0000-0002-3824-5017
Chérif F Matta https://orcid.org/0000-0001-8397-5353
Sam Patrick https://orcid.org/0000-0001-9239-6310
Suzy Lidström https://orcid.org/0000-0003-0050-413X